# An RNA-Based Theory of Natural Universal Computation

Hessameddin Akhlaghpour[ǂ]

[ǂ] Laboratory of Integrative Brain Function, The Rockefeller University, New York, NY, 10065, USA

**Abstract:** Life is confronted with computation problems in a variety of domains including animal behavior, single-cell behavior, and embryonic development. Yet we currently do not know of a naturally existing biological system that is capable of universal computation, i.e., Turing-equivalent in scope. Generic finite-dimensional dynamical systems (which encompass most models of neural networks, intracellular signaling cascades, and gene regulatory networks) fall short of universal computation, but are assumed to be capable of explaining cognition and development. I present a class of models that bridge two concepts from distant fields: combinatory logic (or, equivalently, lambda calculus) and RNA molecular biology. A set of basic RNA editing rules can make it possible to compute any computable function with identical algorithmic complexity to that of Turing machines. The models do not assume extraordinarily complex molecular machinery or any processes that radically differ from what we already know to occur in cells. Distinct independent enzymes can mediate each of the rules and RNA molecules solve the problem of parenthesis matching through their secondary structure. In the most plausible of these models all of the editing rules can be implemented with merely cleavage and ligation operations at fixed positions relative to predefined motifs. This demonstrates that universal computation is well within the reach of molecular biology. It is therefore reasonable to assume that life has evolved – or possibly began with – a universal computer that yet remains to be discovered. The variety of seemingly unrelated computational problems across many scales can potentially be solved using the same RNA-based computation system. Experimental validation of this theory may immensely impact our understanding of memory, cognition, development, disease, evolution, and the early stages of life.

## 1. Introduction

It may be argued that computation is the most fundamental aspect of life. Any problem that involves converting inputs to outputs where the informational content – rather than the material content – defines the problem is a problem of computation. Some examples of computation in biology include: using vision to guide wing movement in insect flight, language acquisition in humans, decision-making in single-celled ciliates [1,2], and embryonic development, the decisional process of beginning with a single cell and coordinating across daughter cells to produce a complex finely-detailed three-dimensional structure. Even though the computations that occur in these settings are poorly understood, it is generally assumed that the mechanistic building blocks that carry out computation in biology have already been identified. In the domain of animal behavior, these building blocks are believed to be neurons and neural networks. And in the domains of cell behavior and embryonic development, they are thought to be molecular cascades, gene regulatory networks, and signal transduction pathways. But the adequacy of these building blocks is disputable and is not rooted in the theory of computation. In fact, if we are to take the theory of computation seriously, it is reasonable to consider that there may exist a computation system that remains undiscovered in biology.

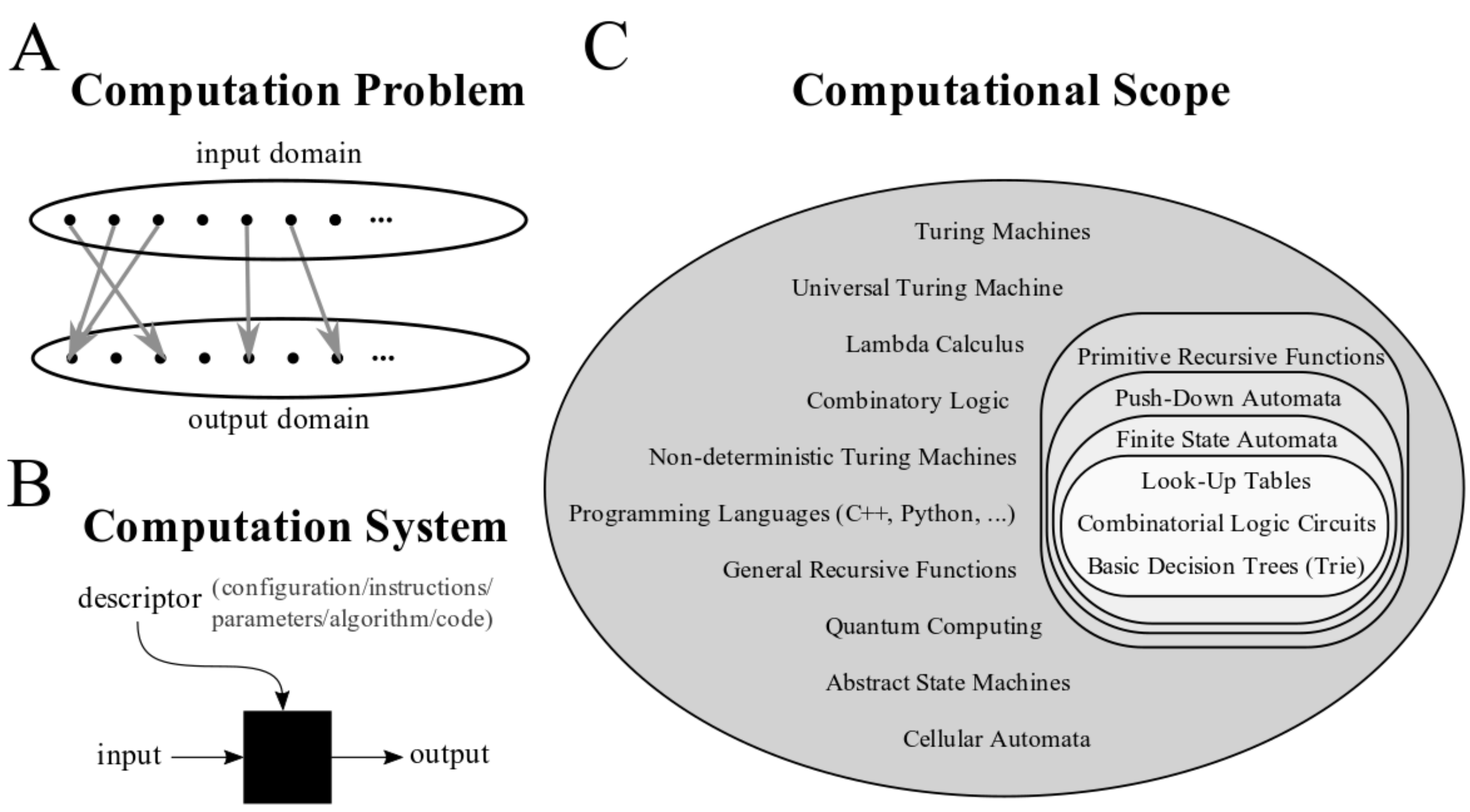

Figure 1: *Basic Concepts from the Theory of Computation* **(A)** A computation problem is a partial mapping between an input domain and an output domain. For instance the problem of squaring a number is a mapping from numbers to their squared values. **(B)** A general-purpose computation system must be able to solve – not just one, but – many problems. Which problem it is to solve is determined by its *descriptor*. The *descriptor* may specify program instructions or the physical configuration that determines how the system works. For example, in a neural network computation system the description is the set of nodes, connections, weights, activation functions, etc. **(C)** A diagram depicting the computational scope of various systems. Computation systems very often emulate one another. The inner-most scope is the weakest form of computation and the outer scope includes all universal (i.e., Turing-equivalent) computers. There are many other scopes not shown in the diagram. These computation systems are all “transducers” rather than “recognizers”; their output domains are strings or natural numbers rather than binary values (i.e., accept/reject).

For every computation system, there are problems that it can solve and problems that it cannot. The set of problems that a system can compute is its *scope*. Measurement of scope is agnostic to how the system works; the components of the system can be analog or digital, discrete or continuous, stochastic or deterministic. Biological computation systems are no exception. They can be analyzed in the framework of the theory of computation (see Fig 1). A system is said to be at least as powerful as another if the former can simulate the latter (i.e., solve all of the problems in the other’s scope). For example, combinatorial logic circuits (i.e., circuits consisting of boolean gates) are equivalent to look-up tables; both systems are capable of solving any problem defined over finite input/output domains. Finite state automata are strictly more powerful than either of those systems and can solve problems that are defined over infinitely large input/output domains (e.g., the problem defined by "given an arbitrarily large number as a string of digits, return the remainder of that number when divided by 7"). But there are some problems that finite state automata cannot solve (e.g., the problem defined by "given a string of open/close parentheses, determine whether it is balanced"). (See [3–5] for a framework for comparing systems that operate on different domains).

One of the most profound mathematical discoveries of the 20th century was that there is a fundamental limit to computational scope [6–11], meaning that there are some definable problems that are uncomputable; no effective computation system with finite means can solve them (see Appendix A for a discussion on what is meant by "effective computation" or "finite means" and why this is relevant to biology). Many abstract systems achieve that scope of computation. Such systems are said to have *universal* scope and are able to compute any computable function. Turing machines are one such system. There are many other universal computation systems, some of which were developed independently and do not resemble Turing machines in any obvious way. All of the universal computation systems listed in Fig 1C are capable of simulating one another, but it is common to define universal computation in terms of Turing-equivalence. A computation system is universal, if and only if it can simulate any Turing machine.

## 1.1 The Universal Computer is Missing in Biology

Finite-dimensional dynamical systems are the dominant computational paradigm in biology, encompassing neural networks, biochemical signaling cascades, and gene regulatory networks. In these models the state of the system is described in terms of a number of physical quantities (such as membrane potentials of different cells or phosphorylation rates of certain proteins, or binding occupancy at various DNA sites). These quantities positively or negatively influence one another. Network models are special cases of dynamical systems where the variables are represented by nodes and the interdependencies are sparse and can be drawn as a network. But physically relevant finite-dimensional dynamical systems are not known to be capable of universal computation. This claim may appear to contradict common wisdom since it was shown in the early 1990s that dynamical systems can simulate Turing machines [12–14]. Based on those results it has been incorrectly asserted that chemical and neural networks are Turing-equivalent [15–17]. But in every instance where a finite-dimensional dynamical system has been shown to be capable of simulating Turing machines, that system lacked *structural stability* [12–14,17–25]. Structural instability renders a system physically unrealizable. It means that even if the system is noiseless and the internal state variables have infinite precision, any arbitrarily small amount of error in the differential equations will lead to a radically different system that does not resemble the intended dynamics. (Structural stability is a distinct concept from chaos; chaotic systems are realizable and physically relevant).

Moore, who was the first to show that finite-dimensional dynamical systems are capable of simulating Turing machines [12,13], argued that *structural stability* is a reasonable criterion for determining whether a dynamical system may be possible to build or found to occur in nature and he conjectured that "finite-dimensional dynamical systems that are structurally stable (or generic by any other reasonable definition) are incapable of universal computation" [26]. Moore's conjecture still stands today and we do not know of any biologically plausible network model capable of universal computation. (See Appendix B for a discussion on computational power of physically relevant dynamical systems).

Another common misconception is that the implementation of a universal logic gate (e.g. NAND or NOR) is sufficient for universal computation (i.e. Turing-equivalence) [27,28]. It is quite easy to construct logic circuits with already identified bio-molecular building blocks [29]. But combinatorial logic (not to be confused with "combinatory logic") has the weakest scope of computation (Fig 1C). Since boolean logic circuits lack memory, the input size is predetermined by the descriptor and they are therefore equivalent to look-up tables in their computations scope. (For example, there is no descriptor for a combinatorial logic circuit that can compute the remainder of any number by 7. But there is a descriptor for a finite state machine that can, and it works for arbitrarily large number).

But why would life need a universal computer? Or what advantages might it have to be favored by natural selection? Perhaps living organisms get by without universal computation and the currently conceived network models are sufficiently powerful to address the problems and opportunities they face. This proposition is difficult to accept given the richness and complexity of life, particularly given how simple it is to achieve universal computation. It is not uncommon to accidentally stumble upon universality in systems where memory usage can expand. Notable examples are Conway's game of life [30], Wolfram's Rule 110 [31], Wang Tiles [32], and Schönfinkel's/Curry's combinatory logic [33,34]. A powerful computing device would immensely benefit organisms that struggle to survive and reproduce. We know evolution is capable of designing remarkably sophisticated systems according to principles of optics, mechanics, chemistry, and thermodynamics. So why not principles of computation? There is nothing about universal computers that make them generally more costly, less efficient, or harder to maintain in comparison to weaker computation systems. In fact, the history of human technology suggests quite the opposite; analog non-universal systems are increasingly being replaced by digital microprocessors in devices and machines even though they do not strictly need universal computation for their purposes. (Microprocessors implement the von Neumann architecture and are a universal computation system).

Let us entertain the possibility that a universal computer exists in biology but has not yet been found. The most obvious principle that may guide us towards finding such a system is *memory expansion*. A necessary [but not sufficient] condition for universality is that memory usage (in systems where "memory usage" can be defined) not be bounded by the system's descriptor (see Fig 1B). In other words, the system should be able to recruit more memory space when needed *during* the process of computation or *after* it is given an input. (Memory usage in finite state automata, a weaker non-universal computation system, is bounded by the system's descriptor irrespective of the input). In the context of neuroscience and cognition, the importance of separating memory from computation and the need for a read-write mechanism has been previously raised by Gallistel & King as a critique of the network paradigm [35].

Network models cannot easily be reconciled with the memory expansion condition. The solution adopted in the Turing-equivalent dynamical systems discussed above is to expand memory by using the numerical digits of a variable as a string of symbols [12–14]. This method has a severe practical limitation; realistically, less than a few dozen bits of memory can be recruited and, more importantly, it leads to structural instability which renders it impossible to build or find in nature. Another solution is to assume that the network can grow in its number of variables. This can be achieved by either having the system physically grow through the construction of new components during computation (e.g., by the generation of new neurons or creation of entirely new molecules/genes according to specific rules) or by assuming there is an arbitrarily large reservoir of *silent* or *dormant* dimensions (e.g., implemented by repetitive network architectures) that serve as general purpose units of memory and can be accessed for storage and retrieval. Crucially, the number of memory units should not be part of the system's *descriptor* or differential equations that determine the problem that is supposed to be solved.

There has been notable progress in recent years toward building network models that use architectural motifs, instead of using numerical digits of a physical quantity as memory symbols. Papadimitriou et al. 2020 implemented the memory tape component of Turing machines using neuronal assemblies, but an exogenous agent was needed to carry out the logic of the Turing machine's tape head [36]. Graves et al. 2014 implemented the Turing machine's tape head as a neural network but their memory bank was a not implemented with a neural network [37]. It remains to be shown whether a fully neuronal model can combine all the components of a Turing machine together in a biologically plausible manner.

Another potentially promising direction is to search for implementations of cellular automata in biology [38]. Cellular automata are potentially universal in scope; depending on the transition rules, it may be possible to use them to compute any computable function (although they may incur slow-downs relative to Turing machines). For instance, Oku & Aihara 2010 modeled a neural network implementation of Rule 110, a Turing-equivalent linear cellular automata [39]. (What can make this model different from standard neural network models is that the size of the network is not bounded by the descriptor; the descriptor can be defined to only specify the initial activity pattern without specifying the full network that that pattern is embedded in). Similarly, Bandini et al. 2005 modeled a reaction-diffusion system that resembled a linear cellular automata capable of simulating Turing machines [40]. Consequentially, reaction-diffusion systems have been suggested as a candidate for computational framework for studying living organisms [41]. A natural implementation of cellular automata presupposes repeating motifs (e.g. biological cells, or repeating network architectures) which contain activity patterns that are invariant to spatial shifts; the evolution of an activity pattern in time should not depend on where that pattern is located in the substrate. This is a crucial property of cellular automata that fulfills the memory expansion condition. A pattern can recruit more memory units by expanding in the surrounding space and every computational program can be described independent of the physically available memory capacity. It is yet to be determined whether cellular automata bear any utility in understanding how biology might achieve universal computation.

Perhaps the most promising place to search for a universal computer is in the molecular biology of polynucleotides. Memory expansion is trivial in a system that uses the precise sequence composition of polynucleotides as memory. It can be accomplished through the addition of nucleotides, either by insertion or tail extension. The resemblance of polynucleotides to strings of computation theory is hard to ignore. In Turing's attempt to formalize the notion of computation he wrote: "*Computation is normally done by writing certain symbols on papers... I think that it will be agreed that the two dimensional character of paper is no essential of computation. I assume then that the computation is carried out on a one-dimensional paper, i.e., on a tape divided into squares. I shall also suppose that the number of symbols which may be printed is finite... The effect of this restriction on the number of symbols is not very serious. It is always possible to use sequences of symbols in the place of a single symbol*" [9].

Polymer sequences consisting of an alphabet of only four nucleotide symbols A, C, T (or U), and G elegantly fit this description. This striking resemblance was first noticed in the 1970s by Bennett who later proposed a blueprint for a RNA/DNA based Turing machine [42,43]. Apart from their string-like structure, polynucleotides possess many properties that make them ideal vehicles for biological computation, e.g., thermodynamic stability, spatial compactness, and their capacity to be modified with low energy cost [44,45]. It is not surprising that these molecules are considered to be potentially useful in data storage technology [46,47] and biomolecular computing [29,48,49]. DNA/RNA computation has been used in laboratory settings to solve a number of challenging tasks including the Hamiltonian path problem [50], the travel salesman problem [51], the boolean satisfiability problem [52–54], and the knight placement problem [55].

Already, a number of DNA/RNA-based computation models have been proposed that are universal in their computational scope. One approach is to have DNA strands self-assemble into two-dimensional structures that obey the laws of Wang tiles or linear cellular automata [56–59]. In this approach, memory expansion is achieved through the growth of an aperiodic crystal. Another general approach is to use DNA strand displacement to build a chemical reaction network [60–63]. It has been suggested that chemical reaction networks can be used to implement counter machines (which are Turing-equivalent) by relying on the exact number of molecules that are present [62]. A third theoretical approach is to use DNA strand displacement to implement stack machines or Turing machines [64–66], where memory expansion is achieved by the lengthening of the DNA strand at its ends. A fourth approach is to enzymatically modify the content of a DNA strand that serves as a Turing machine memory tape [42,67–69]. A more recent approach is to exploit the complexity of folding pattens of RNA molecules during transcription [70,71]. Rothemund's approach, which may be the earliest detailed Turing-equivalent model, was based on restriction enzymes acting on a circular double stranded DNA, where complementary DNA overhangs are ligated [72].

Although one of the goals of DNA/RNA based computation has been to implement it in living cells [73–78], it would be hard to make the case that any of these Turing-equivalent models already exist in nature as a general purpose computation system. The above body of work has been design-oriented, not discovery-oriented. The question I would like to raise here is not whether it is feasible to artificially implement a universal computation system using polynucleotides, but whether it is reasonable to consider that nature may have already done so. This question will lead us to an RNA-based model that looks very different from the ones that have been proposed to date.

Recent developments in molecular biology suggest the possibility that non-protein-coding RNA have a yet undiscovered critical role. Approximately 1.74% of the human genome ends up in mature mRNA and more than half of that number is untranslated regions that do not encode proteins [79]. This is while it is has been found that the vast majority of the human genome is actively transcribed [80–82]. The discovery of pervasive transcription has been met with controversy [81–87] and the functional significance of the non-coding portion of the transcriptome is being intensely debated [87–97]. Proponents of the "junk DNA" hypothesis estimate that no more than 15% of the human genome can have functional significance, and the rest leads to "transcriptional noise" when transcribed [98–100]. The opposing viewpoint argues that most of the human genome may be functional and that sequence conservation is not a necessary condition for functional relevance [92,94,97,101] and that there are many other indicators of function such as conservation of secondary structure [102,103], conservation of promoters sequences [104–108], cell-specificity in expression levels [104,109–111], subcellular organization [112,113], and temporal regulation during embryonic development [114,115]. Genome-wide association studies show that more than 70% of the genetic loci associated with traits and diseases fall in intergenic or intronic regions [90]. These regions have been found to be abundantly transcribed [116,117] in a highly cell-type specific manner consistent with their associated traits [118]. Across organisms the non-protein-coding to protein-coding ratio of the genome scales with organism complexity, while the number of protein-coding genes as well as the total length of protein-coding sequences plateaus [119–121]. While there are many indicators suggestive of function, the mechanistic roles of non-coding RNA remain to be discovered. The question of non-coding RNA function has even been described as "*the most important issue in genetics*" [81]. I propose the theory that the non-protein-coding portion of genome and transcriptome contains the data and programming material of an undiscovered universal computation system in biology.

## 2. Natural RNA-Based Universal Computation is Plausible

In support of the theory of natural universal computation through polynucleotides I demonstrate that universal computation through RNA is in principle attainable without assuming extraordinarily complex molecular machinery. A molecular machine that implements a universal Turing machine would be extraordinary and implausible, as it would require large enzymes operating in a far more elaborate manner than the ribosome. Even if such a system were molecularly feasible, it is hard to imagine how it could have gone undetected. I present an alternative class of models based on λ-calculus and combinatory logic. Computation in these models is a decentralized process where distinct enzymes make local modifications to RNA molecules according to basic rules. The specific details of the models are somewhat arbitrary and only meant to be used as a proof a principle, demonstrating that it is possible to implement a universal computation system through basic molecular operations on RNA. I argue that the models are plausible and that it is conceivable such a system may have evaded detection throughout the many decades of research in molecular biology.

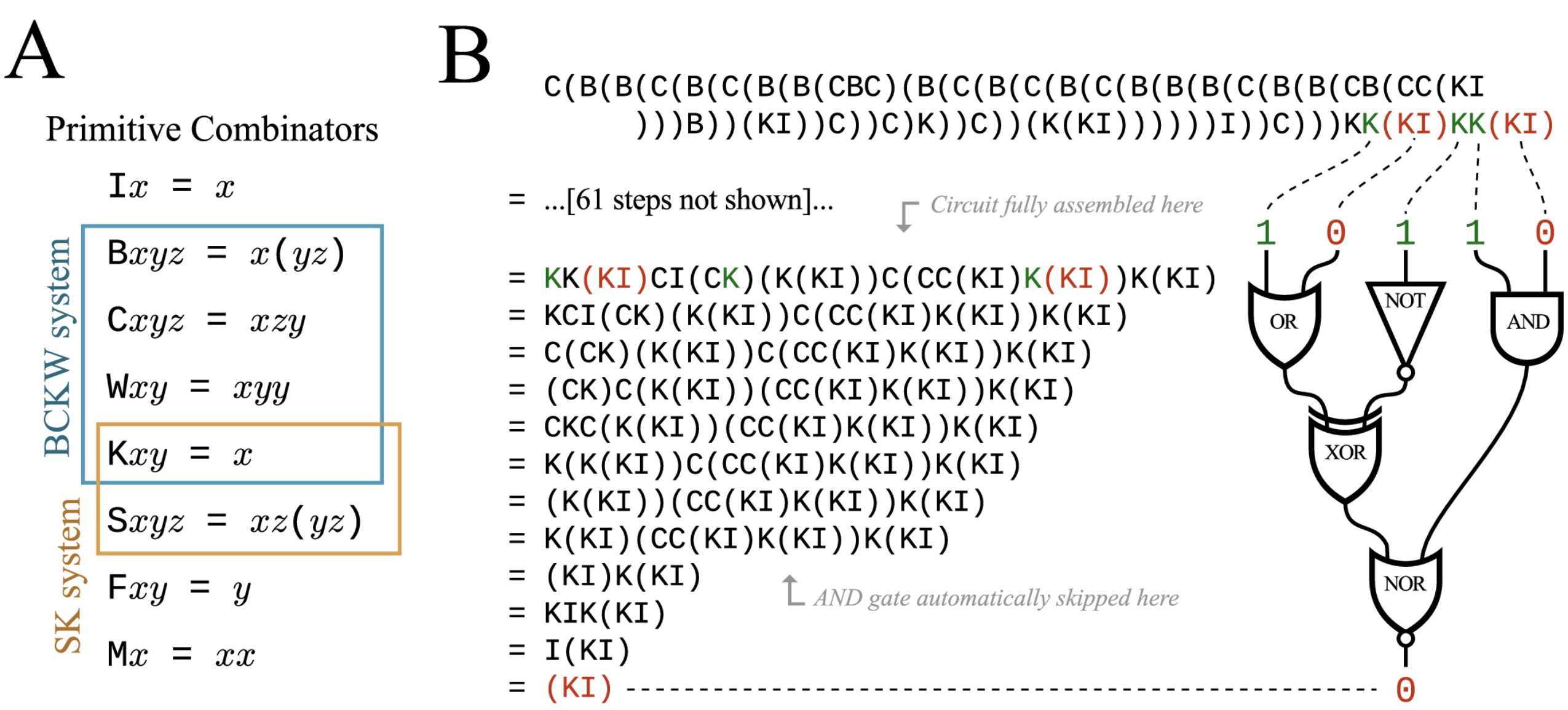

Figure 2: *Implementation of a combinatorial logic circuit using combinatory logic* – **(A)** Definitions of some commonly used primitive combinators. Outlined are two common universal systems: SK and BCKW. **(B)** A combinatory term was constructed using combinators B, C, K, and I to compute a circuit with five boolean inputs. Here, Church's boolean encoding is used, where K represents *true* and KI represents *false*. At each step, either the parentheses enclosing the left-most term are deleted or the left-most combinator is applied. The 61 omitted steps only rearrange the five inputs and assemble the circuit. Evaluation of the AND gate was automatically skipped since the first input to the NOR gate was true. The final result is KI, equivalent to false. Combinatory logic must not be confused with combinatorial logic. The latter is the weakest of the systems in Fig 1C (equivalent to look-up tables), whereas the former is a universal computer (equivalent to Turing machines).

## 2.1. Combinatory Logic and λ-calculus as Computation Systems

λ-calculus and its predecessor, combinatory logic (CL), are two nearly identical universal computation systems. The entities defined in these systems are functions that take functions and return functions. No distinction is made between programs and data and everything is constructed as a function. There is a one-to-one equivalence between lambda functions (called "λ-terms") and combinatory functions (called "combinators"). The difference lies in the elementary operations that are used to compute things. λ-calculus computes using variable substitution and variable renaming. CL uses applications of primitive combinators. (I will only briefly introduce CL here. For a more complete introduction to both systems see Appendix C).

The identity combinator **I** is defined as **I***x* = *x*. It returns whatever it is given. The combinator **K** is defined as **K***xy* = *x*. It takes two arguments and returns the first. **C** takes three inputs and swaps the second and third, **C***xyz* = *xzy*. **B** is defined as **B***xyz* = *x*(*yz*), **W** is defined as **W***xy* = *xyy* (see Fig 2A for more primitive combinators). (Combinators are capitalized and variables are in lower-case). Let us evaluate the term **CIBW**. At every step we apply the left-most combinator. Applying **C** results in **IWB**. Next we apply **I** to get **WB**, which cannot be evaluated further because application of **W** requires two inputs. So the final result is **WB**. Let us try another example, this time with a variable: **CKC***x* = **K***x***C** = *x*. Like the **I** combinator, **CKC** returns *x* given any *x*. This shows that **I** can be constructed using **C** and **K**. Finally, let us evaluate an example with parentheses: **BCKIW(KK)KC**. First we apply **B** to get **C(KI)W(KK)KC**. To apply **C** we swap the second and third arguments, **W** and **(KK)**, to get **(KI)(KK)WKC**. We can always remove the parentheses around the left-most term because CL is by convention left-associative. Doing so, we get **KI(KK)WKC**. Applying **K** we get **IWKC** = **WKC** = **KCC** = **C**. So **BCKIW(KK)KC** = **C**.

Remarkably, the set of primitive combinators **B**, **C**, **K**, and **W**, constitute a universal computation system. Using only these four combinators, it is possible to simulate Turing machines and compute any computable function. Not only is it possible to compute boolean logic circuits (see Fig 2B), but CL can implement data structures and recursive algorithms (see Appendix C). For example **C**(**C**(**B**(**BK**))*z*)(**C**(**C**(**B**(**BK**))*y*)(**C**(**C**(**B**(**BK**))*x*)(**KI**))) can be interpreted as a stack containing three elements x, y, and z. And **B**(**WI**)(**BWB**)(**B**(**B**(**C**(**B**(**B**(**B**(**C**(**BC**)**I**)(**BC**(**CI**))) (**BC**)))(**BC**(**C**(**B**(**BK**)))))**B**)**B**)(**KI**) is a recursive program that takes a stack of any size and reverses the order of its elements. (To see how I constructed these terms see Appendix C). CL can implement numbers and arithmetics. The most commonly used number systems are unary (e.g., Church numerals). This has given λ-calculus and CL a reputation for being slow. But it is not difficult to implement efficient arithmetic operations with binary numeral or in any base of choice [122]. λ-calculus and CL are as powerful and expressive as any functional programming language, and can simulate Turing machines with linear slow-down (see Appendix C for proof).

## 2.2. Combinatory logic can be implemented through RNA Editing Rules

In CL, terms are usually represented as strings of characters. Nucleotide sequences can trivially be used to represent strings. There is already a precedent for this in protein coding sequences where each triplet represents an amino-acid. Similarly, each primitive combinator can be encoded using sequence motifs. (Open and close parentheses can also be represented by unique motifs, but this is not the method used in the models I present below.) Not all nucleotide sequences must represent a combinator (some can be neutral fillers) and combinator motifs need not be unique (there may be redundancy similar to amino-acid codons). A small enough primitive combinator set, like S and K, can make it possible to use just one nucleotide per combinator. Non-canonical bases and nucleotides modifications like methylation may also be involved in the representation scheme. I refrain from speculating over the motifs for the combinators, but I only remark that if the codes for the primitive combinators are all of the same length, it can make molecular implementation of the combinator rules simpler.

To evaluate a term, it is sufficient to recursively apply the leftmost combinator, as illustrated in the examples above, until it is no longer possible to reduce the term. This can be accomplished by distinct and independent enzymes, each responsible for implementing one of the primitive combinators. (The enzymes need not be proteins; they can be other RNA strands or even self-cleaving/self-ligating RNA elements). For example, if there are four primitive combinators there can be four distinct and independent enzymes that each apply one of the combinators by first recognizing the left-most motif that encodes for it and then applying appropriate changes to the RNA strand. These applications can be carried out in an uncoordinated fashion and there is no need for a central entity to direct these operations.

If we take nucleotide sequences to represent CL terms, it seems more reasonable to assume that they are parsed and edited at the 3'-end. The 5'-end is typically capped and protected whereas the 3'-end is highly dynamic and amenable to additions or removals of nucleotides; RNA synthesis occurs by polymerization at the 3'-end and the poly(A) tail (a stretch of up to 200 adenine bases at the 3'-end) can lengthen or shorten even after an RNA is exported to the cytoplasm [123,124]. Additionally, RNAs exhibit high diversity in their 3' splice site. The same RNA transcript can be cleaved and polyadenylated at different sites depending on the context [125,126]. The variability of the 3'-end is especially pronounced in the brain where 3'UTRs of mRNA are lengthened beyond the annotated ends [127,128] and where miRNA show extensive sequence modifications at their 3'-end [129]. For these reasons I will assume that RNA strands are parsed as combinatory terms in the 3'-to-5' direction. Respecting the standard nomenclature, I refer to the direction of the 5'-end as "upstream" and the direction of the 3'-end as "downstream".

What sort of RNA modifications are required to implement combinator applications? The answer depends on the primitive combinator set. **B**, **C**, **K**, and **W** are sufficient for universality. So is the smaller set of only two combinators **S** and **K**. (All four combinators **B**, **C**, **K**, and **W**, can be constructed using only **S** and **K** – see Appendix C). There are infinitely many valid basis sets that would produce a universal computation system. Fortunately, we can evaluate the plausibility of the model without assuming the exact basis set. For any set of primitive combinators to achieve universal computation, it must have at least one combinator that deletes terms (**deletion**), at least one that reorders terms (**reordering**), at least one that duplicates terms (**duplication**), and at least one that adds parentheses (**nesting**). (This is easy to prove by showing that none of these operations can be mediated by the other three. Without deletion the number of combinators cannot shrink. Without duplication the number of combinators cannot grow. Without reordering the ordering of terms remains invariant. And without nesting the number of parentheses – in, say, left-associative representations – cannot grow). In the **BCKW** system, each of the four combinators fulfills exactly one of the four conditions. In the **SK** system, the **S** combinator fulfills the last three conditions and the **K** combinator fulfills the first. We can examine the plausibility of each of these four operations separately.

**Deletion:** The enzyme responsible for implementing the deletion operation needs to excise a segment of RNA. Splicing is one of the most common RNA modifications and many enzymes are known to mediate it. Splicing involves cleavage and ligation but even a single cleavage operation is sufficient to completely fulfill the deletion condition. For instance, the **F** combinator, equivalent to Church boolean *false*, is a combinator that takes two inputs and returns the second (**F***xy*=*y*). An enzyme responsible for implementing **F** only needs to cleave the RNA one term upstream of the **F** motif (corresponding to deletion of "Fx" in "Fxy"). (The **K** combinator, defined as **K***xy* = *x*, can be constructed as **K**=**CF** and no longer needs to be in the basis set). RNA strands that are excised through this method must be immediately discarded and not interpreted as representing combinatory terms. This can be trivially implemented since excised strands lack a 5'-cap and are susceptible to exonuclease degradation.

**Reordering:** The enzyme responsible for implementing the reordering operation only needs to conduct something as simple as swapping two elements corresponding to successive terms, e.g., as in **C***xyz*=*xzy*. Similar RNA modifications are already known to occur in cells. Post-transcriptional re-ordering of exons was first observed in the early 1990s [130,131]. At first, it was thought to be rare, expressed at low levels, and confined to circular RNAs. But several recent studies suggest that it may occur abundantly, occurring in polyadenlyated transcripts at expression levels comparable to that of their canonically spliced counterparts [132–137]. Transposable elements in DNA, RNA's sister molecule, frequently move around changing locations with the help of transposase enzymes that mediate their cleavage and ligation. Operationally it only requires three cleavages and three ligations to fulfill the reordering condition, both of which many native enzymes like the spliceosome are capable of [132]. It is therefore plausible to assume other enzymes may exist that are capable of reordering RNA elements.

**Duplication:** Duplication is not as trivial as deletion or reordering. Example combinators that require duplication are are **W***xy*=*xyy* and **M***x*=*xx*. The term that needs to be duplicated may either be a motif for a single combinator or a nested (i.e., parenthesized) sequence of arbitrary length. Duplication of arbitrarily long sequences requires an enzyme that can synthesize a copy of an RNA element. RNA dependent RNA polymerases (RdRp), the enzymes that can directly synthesize RNA from an RNA template, are common in viruses and have also been found in plants and nematodes. But many species including humans and fruit flies lack endogenous RdRp. Are there any known methods of RNA sequence duplication that may exist in all cells? One method is through reverse transcription (i.e., DNA synthesis from an RNA template) followed by transcription, (RNA synthesis from a DNA template). Reverse transcriptase and RNA polymerase, exist abundantly across the plant and animal kingdom, although, most bacteria species lack reverse transcriptase [138,139] and reverse transcriptase is thought to be inactive in many – if not most - cells of multicellular eukaryotes.

But a more promising candidate enzyme for mediating sequence duplication in the model is RNA polymerase (RNAP). RNAP is the crucial transcription enzyme, abundantly present in all living organisms, that normally synthesizes RNA from DNA templates. In special cases, RNAP can polymerize RNA solely from RNA templates [140]. RNA replication through RNAP is the method used by viroids that infect plants, the hepatitis delta virus (HDV) that infects humans, and several other related RNA based viruses that lack their own polymerase enzyme [141–144]. These viruses rely on native RNAP in host cells for RNA replication. In the case of viroids and HDV, the replicated RNA is circular and replication is understood to happen through a rolling circle mechanism [144,145]. RNA replication of non-circular RNA strands has been demonstrated in vitro [146–152] Remarkably, this mode of replication is dependent on the existence of a 3'-GG... or 3'-CC... motif on the template, begins synthesis immediately after the motif, and can generate concatemers from linear (non-circular) templates [148,150]. This is already quite close to what would be expected of an enzyme responsible for implementing the **W** or **M** combinators. (It must first recognize the motif that codes for **W** or **M**, begin synthesis immediately upstream of the motif, and produce a strand with two copies of the term that is intended for duplication). At any rate, the hypothetical duplication enzyme may still be unknown to us and there is considerable evidence that direct RNA duplication through unknown mechanisms occurs endogenously [133,153–156]. (In the next section we will revisit the problem of RNA duplication and show that it can be outsourced to DNA-based RNA transcription).

**Nesting:** A nesting operation composes two terms as a single nested term. For example, **B** defined as, $\mathbf{B}xyz=x(yz)$, nests its second and third arguments as a single term ($xyz$ is interpreted as $((xy)z)$ in left-associative CL). If parentheses are encoded by sequence motifs, as suggested earlier, the enzyme responsible for implementing **B** needs to insert predefined open and close motifs one and three terms upstream of the motif for **B**. But this method brings out a complication that we have ignored until now: parenthesis matching.

Parenthesis matching is critical for the model because all of the enzymes that apply combinators need to recognize and count whole terms. For example, if the hypothetical deletion enzyme is confronted with an open parenthesis at the position of a term that is supposed to be deleted, it must delete the entire stretch of nucleotides between that open parenthesis and its matching close parenthesis (which can include other open/close parentheses). How can an enzyme when confronted with the start of a term find the correct ending nucleotide? This is not a simple task. A parsing algorithm that implements parenthesis matching needs to keep track of the parenthesis depth, incrementing with every "(", decrementing with every ")", and stopping whenever the calculated depth reaches zero. A molecular implementation of this algorithm does not appear plausible; I cannot conceive of an implementation of this algorithm without invoking extraordinarily complex hypothetical molecular machinery. Fortunately, there is an elegant solution to this problem using the base pairing properties of RNA.

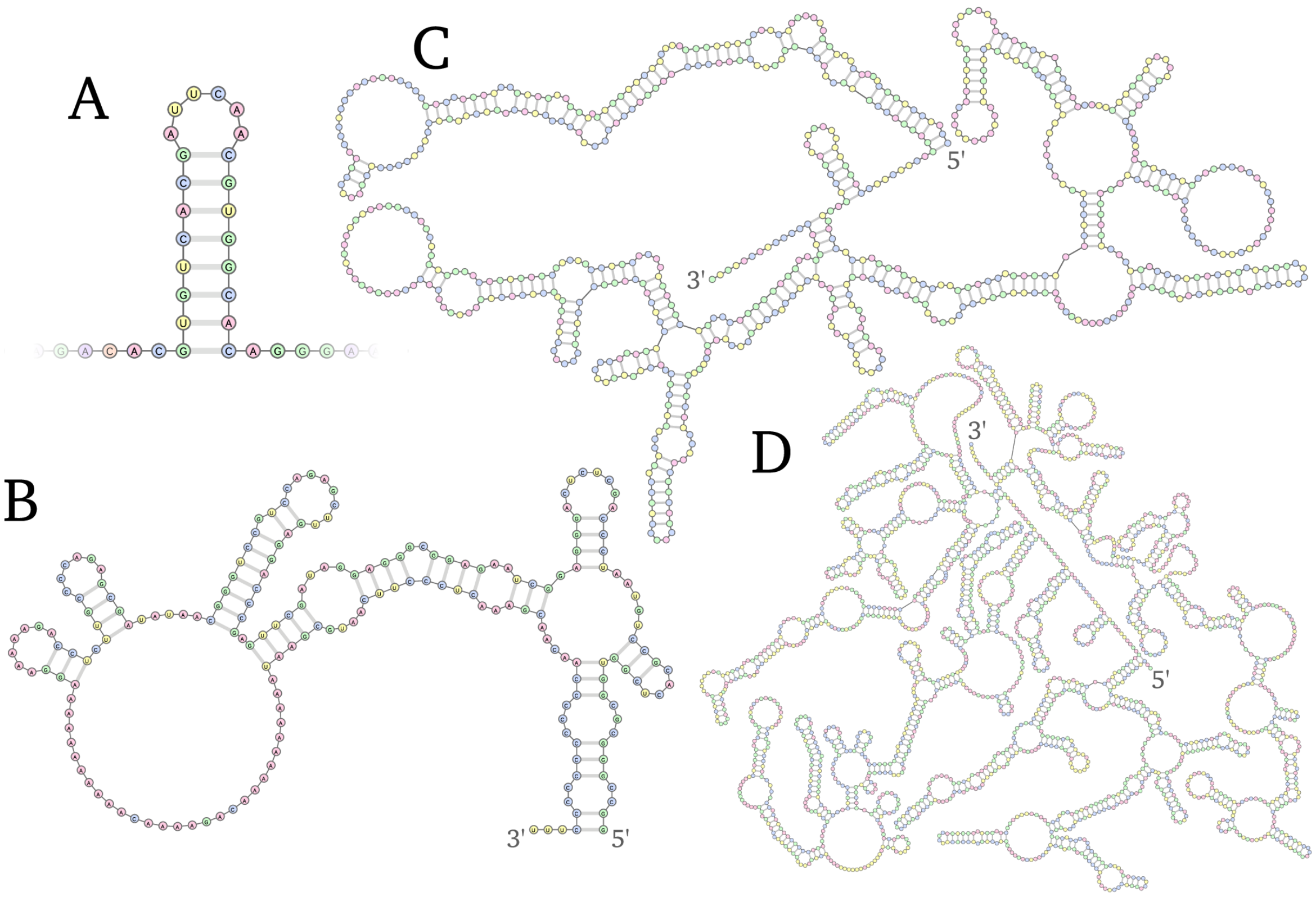

Figure 3: *RNA Secondary Structure* – **(A)** RNA stem-loops form when the nucleotides of two segments of an RNA strand pair with one another. The region containing base-pairs is the stem and region in between the stem is the loop. Base pairing in RNA follows the canonical Watson-Crick rules where A pairs with U and G pairs with C but can also include less stable wobble pairs such as G-U. **(B-D)** Examples of secondary structure of some non-protein-coding RNAs exhibiting many levels of nested stem loops. **(B)** 200-nt BCYRN1 transcript, expressed in neuronal dendrites and implicated in memory loss and cancer [269]. Secondary structure obtained using RNAfold minimum free energy prediction [165,270,271]. **(C)** 683-nt lncTCF7 transcript, implicated in cancer. Secondary structure obtained from [272]. **(D)** 2,148-nt HOTAIR transcript, expressed in peripheral tissues and implicated in epidermal differentiation and development. Secondary structure obtained from [273]. Figures drawn using VARNA [274].

Similar to DNA, RNA molecules can form double stranded helices with their complementary sequences. Base pairing can even occur within the same strand. When an RNA molecule contains two sequences that are inverse complements of one another, those sequences physically come together to form a stem-loop (Fig 3A). Stem-loops can occur inside other stem loops and the entire base-pairing organization of an RNA molecule is referred to as the secondary structure. RNA strands typically have intricate secondary structures that involve many layers of nested stem loops (Fig 3B-D).

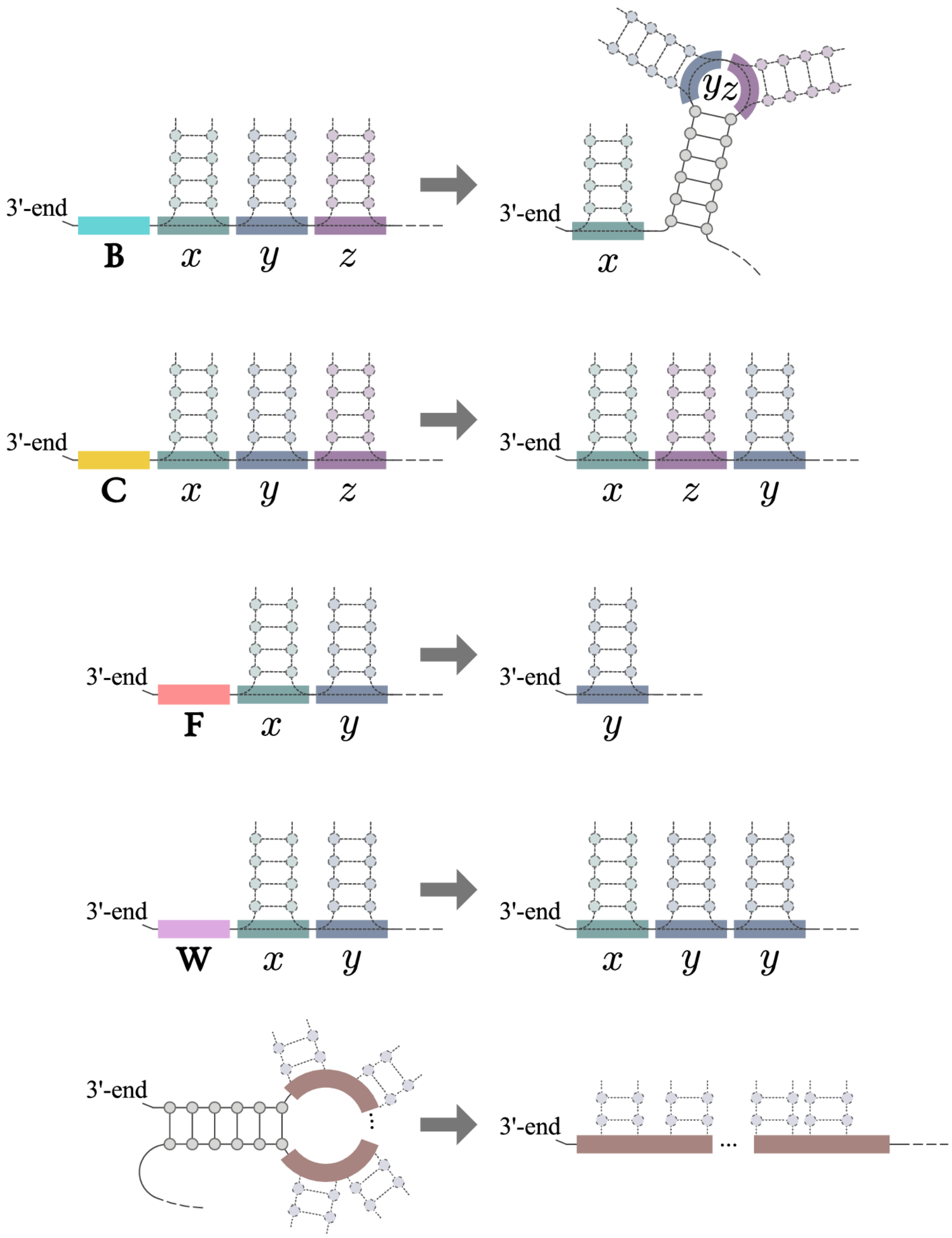

Figure 4: *Model I* - A set of RNA editing rules that mimic the operational rules of left-associative CL based on combinators **B**, **C**, **F** and **W**. In all the rules, RNA elements x, y, and z must either be single combinator motifs (codes that represent **B**, **C**, **F** or **W**) or stem loop bases. The fifth rule corresponds to parenthesis removal when the left most term is enclosed in parentheses. All edits are designed to be applied on the 3'-end of RNA strands.

If open and close parentheses are represented by reverse complementary sequences, RNA molecules naturally solve the problem of parenthesis matching by physically bringing matching parentheses together in space. It is then enough for the combinator enzymes to treat the base of a stem as a single term, just as they would for a primitive combinator motif. For example, a hypothetical enzyme that implements $\mathbf{F}xy = y$ must cleave one term upstream of the **F** motif. If a stem loop appears in the place of $x$, it can delete the entire stem loop by cleaving at the base of the stem. This model suggests a very general role for RNA secondary structure that is more fundamental than anything presently conceived [157].

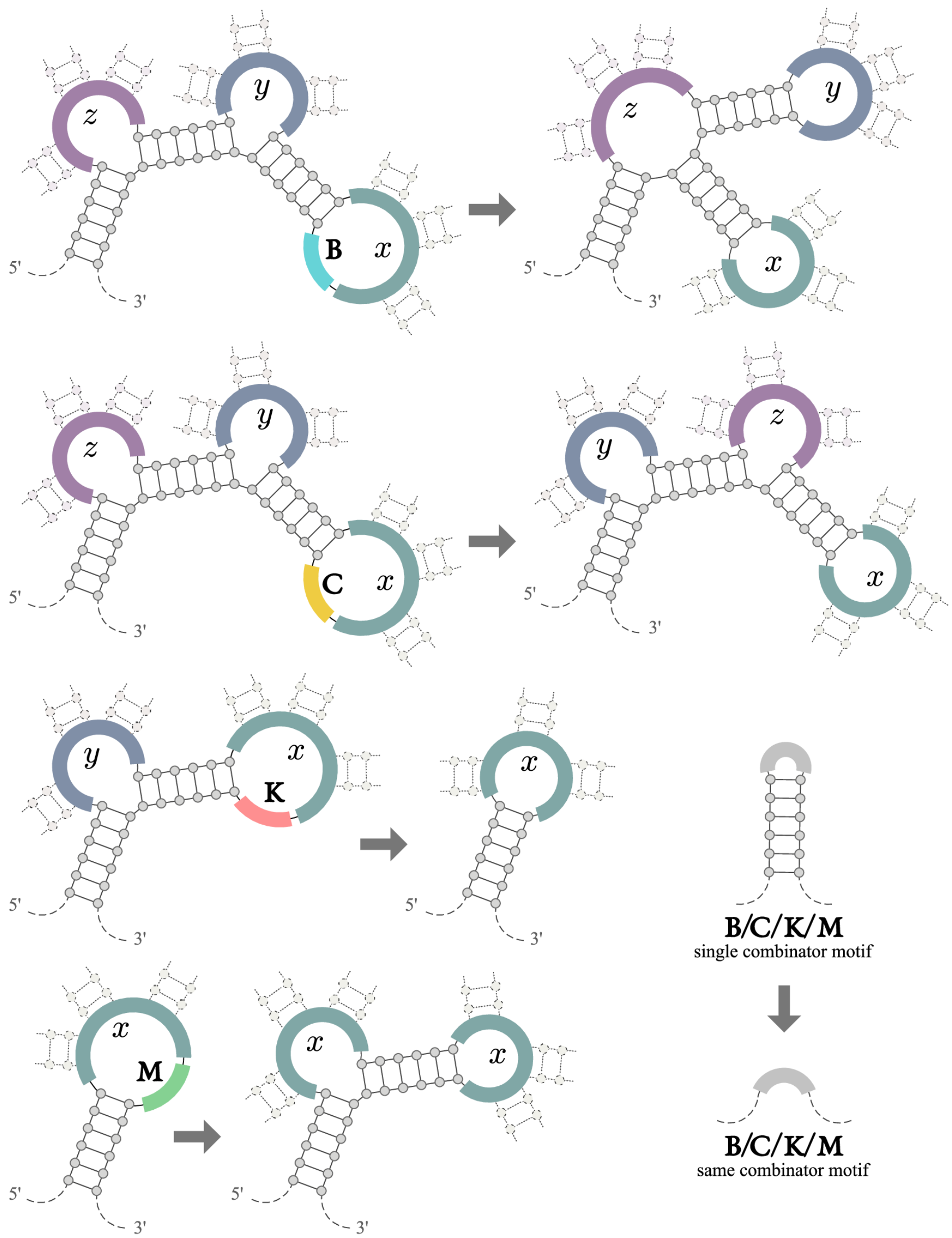

Figure 5: *Model II* - A set of RNA editing rules that mimic the operational rules of right-associative CL based on combinators **B**, **C**, **F** and **K**. In all the rules, RNA elements x, y, and z can be any non-empty sequence representing primitive or composed combinator terms. The fifth rule corresponds to parenthesis removal when a single combinator is enclosed in parentheses. In contrast to Model I (Fig 4) editing can happen anywhere along an RNA strand, even multiple locations at once. If multiple locations can be edited, the order of operations does not affect the final result (according to the Church-Rosser theorem).

In light of this solution, let us evaluate the plausibility of a nesting operation. To implement nesting, there must be a method of adding new stem loops into RNA strands. This can be done through insertion of RNA duplexes. Double stranded RNA (dsRNA) is known to exist in cells and its over-expression or under-expression can be lethal [158,159]. The enzyme responsible for implementing nesting may recruit these dsRNAs, or possibly recycle duplexes that have been removed in previous CL operations, and insert them in place. (A basic operation that the model needs is parenthesis removal. If the left-most term is enclosed in parenthesis, the parentheses should be removed. Removed duplexes can be reused in the same strand for nesting operations). Insertion of RNA duplexes has not been documented in cells, but it would only involve simple cleavage and ligation operations.

Now that we take stems to represent parentheses, it is possible to provide a set of concrete RNA editing rules that can, in theory, implement combinatory logic. The five rules depicted in Fig 4 implement left-associative CL based on four arbitrarily chosen combinators. In left-associative CL *abcde* is interpretted as $(((ab)c)d)e$. A variant of these rules can be constructed by mimicking right-associative CL (see Fig 5) where *abcde* is interpreted as $a(b(c(de)))$. In right-associative CL, the operation rules can be implemented anywhere along the RNA strand, whereas the rules written for left-associative CL were designed to be implemented at the 3'-end of the strand only. It is sufficient for these editing rules to be implemented by distinct enzymes in an uncoordinated fashion to achieve a universal computation system.

The method of nesting terms through RNA stem loops presents an opportunity to implement addressable memory and variable substitution. One such implementation is illustrated in Fig 6. In this model, each variable is assigned an *address* (specified by a unique sequence of nucleotides). An RNA strand containing that address sequence at its 5'-end stores the value of that variable. A variable can be referenced to by other RNA strands using a *reference* sequence defined as the reverse complementary sequence of the address sequence. Simple cleavage/ligation operations can substitute the value of the variable as a nested term as shown in Fig 6. (For this system to work with the left-associative CL model (Fig 4), the address and reference sequences must themselves fold into stem loops).

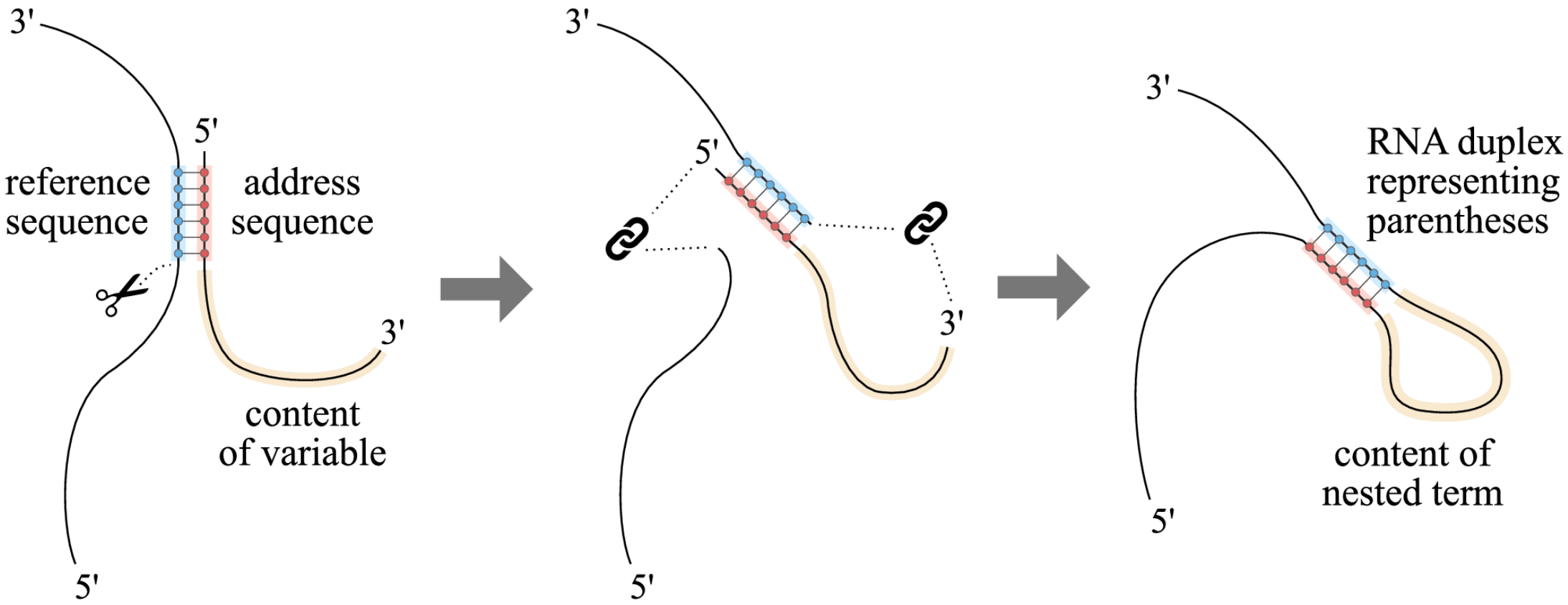

Figure 6: *Addressable Memory* - An example mechanism that can implement variable substitutions through RNA, assuming stem loops hold nested terms. First two distinct RNA strands come together, one containing the address and content of a variable and the other containing a reference sequence that base-pairs with the address sequence. One cleavage and two ligations as illustrated is sufficient incorporate the content of the variable as a stem loop at the location it is referenced. This process in reverse can also be used to excise a loop from a stem-loop, while maintaining the link across two strands.

An addressable memory system with variable substitution has many advantages. First it allows longer strands to be broken down to shorter ones while maintaining a link between the strands. This overcomes some of the physical limits on the lengths of RNA programs and parallelizes computation. Second, it allows coordination across many programs that share the same memory space. A program can write the result of a computation in a designated memory address that another program uses as an input. Third, it can be more efficient to work with variable representations. For example, if a nested term is to be duplicated only to have one copy deleted, it is more efficient to have it represented by a short reference sequence and duplicate the reference sequence. And fourth, term permutations can be done much quicker using addresses and references. The first 61 steps of the program in Fig 2B only rearranges the inputs, but can be accomplished with only 5 variable substitution operations.

## 2.3. Standard DNA-dependent transcription eliminates the need for RNA duplication

Above, we evaluated the plausibility of *deletion*, *reordering*, *duplication*, and *nesting* in RNA molecules. Deletion, reordering, and nesting can be done with O(1) number of cleavage/ligation operations, but duplication requires RNA synthesis and its number of operations is proportional to the length of the element that is being duplicated. Furthermore, RNA replication from RNA templates is not yet known to occur widely across all cells of all lifeforms. In this section I show that RNA directed RNA replication is not strictly required for universal computation. To make this possible I will abandon pure combinatory logic as this model requires an addressing system where self-referencing (or cycles in the reference graph) is permitted.

Strictly speaking, it is not permitted to define a function in terms of itself in λ-calculus or CL. For instance the function ***f*** **= C(C(BC(CC)))(CK)** ***f*** must be redefined in terms of the **Y** combinator as ***f*** **= Y C(C(BC(CC)))(CK) = B(WI)(BWB)C(C(BC(CC)))(CK)** (see Appendix C). (The founders of λ-calculus/CL were concerned with creating a sound foundation for mathematics that avoids contradictions and ill-defined self-referencing terms). But if we relax this constraint it becomes possible to achieve universal computation without any duplicating combinators (such as **W** or **M**). Instead, the three combinators **B**, **C**, and **K** alongside an addressing system is sufficient to simulate any Turing machine with linear slowdown (see Appendix C for a constructive proof that uses B, C, and K, and self-reference to linearly implement any Turing machine). Note that in the model of right-associative CL (Fig 5) applications of B, C, K, and parenthesis removal can be fully accomplished by merely cleaving and ligating RNA at predetermined positions relative to the combinator motif. Except for the application of the **M** combinator (which is no longer needed for universality), the number of duplexes (i.e., parentheses) on both sides of the rules are the same. Therefore, even duplex insertion is not strictly required.

We can combine the addressing system of Fig 6 with the right associative model of Fig 5, excluding the M combinator – which is the combinator that cannot be implemented purely by cleavage and ligation. The resulting model relies on perpetual transcription of RNA strands from static genomic templates. The transcripts can then be recursively inserted into one another guided by an addressing system.

In order to prevent infinite loops of self-insertion where multiple copies of a self-referencing transcript get inserted into one another, the value substitution mechanism of Fig 6 can be reformulated to only occur when a reference is at the left-most position (the 3'-end of the strand) or when it is enclosed in parentheses on both sides. An alternative solution is to split a reference sequence into two subsequences only to have them joined when value substitution is needed in the algorithm.

## 2.4. Simulating An Example Program: Addition of Two Arbitrarily Large Numbers

To better illustrate the idea of computation through RNA, I will explain how to perform arithmetic addition using the RNA model described in section 2.3. Note that the problem of adding two bounded numbers (say two 8 bit numbers, between 0 and 256) can be solved using a static memory-less circuit of boolean logic gates. But our example demonstrates how to use recursion to solve addition with no bounds on the input size, apart from resource constraints like time, space, and energy.

In our example, a numbers is represented as a strand with distinct RNA elements for binary digits 1 and 0. (Examples of the numbers 6 and 2, respectively $(110)_2$ and $(10)_2$ in binary form, are depicted in Fig 7C). Two RNA strands shown in Fig 7A & 7B need to be perpetually transcribed from DNA templates. These two transcripts (hereon referred to as transcripts A and B) have unique address sequences at the 5'-end and can be referred to by the reverse complementary sequence of those address sequences. For instance the RNA strand in Fig 7C has a reference to transcript A near its 3'-end which should lead to transcript A being inserted at that location through the rule illustrated in Fig 6. Additionally, transcript A has a reference to itself and a reference to transcript B. Transcript B also has a reference to itself.

Here, we only use three combinators: B, C, and K. The rules for applying them are shown in Fig 5. Except for the rule for applying the M combinator (which we make no use of), all of the editing rules in Fig 5 can be implemented by only cleavage and ligation at fixed positions relative to the combinator motifs. It is assumed that hypothetical enzymes detect these motifs and collectively implement these rules in any order. To see the full derivation of the addition program, refer to Appendix C section 6.1.

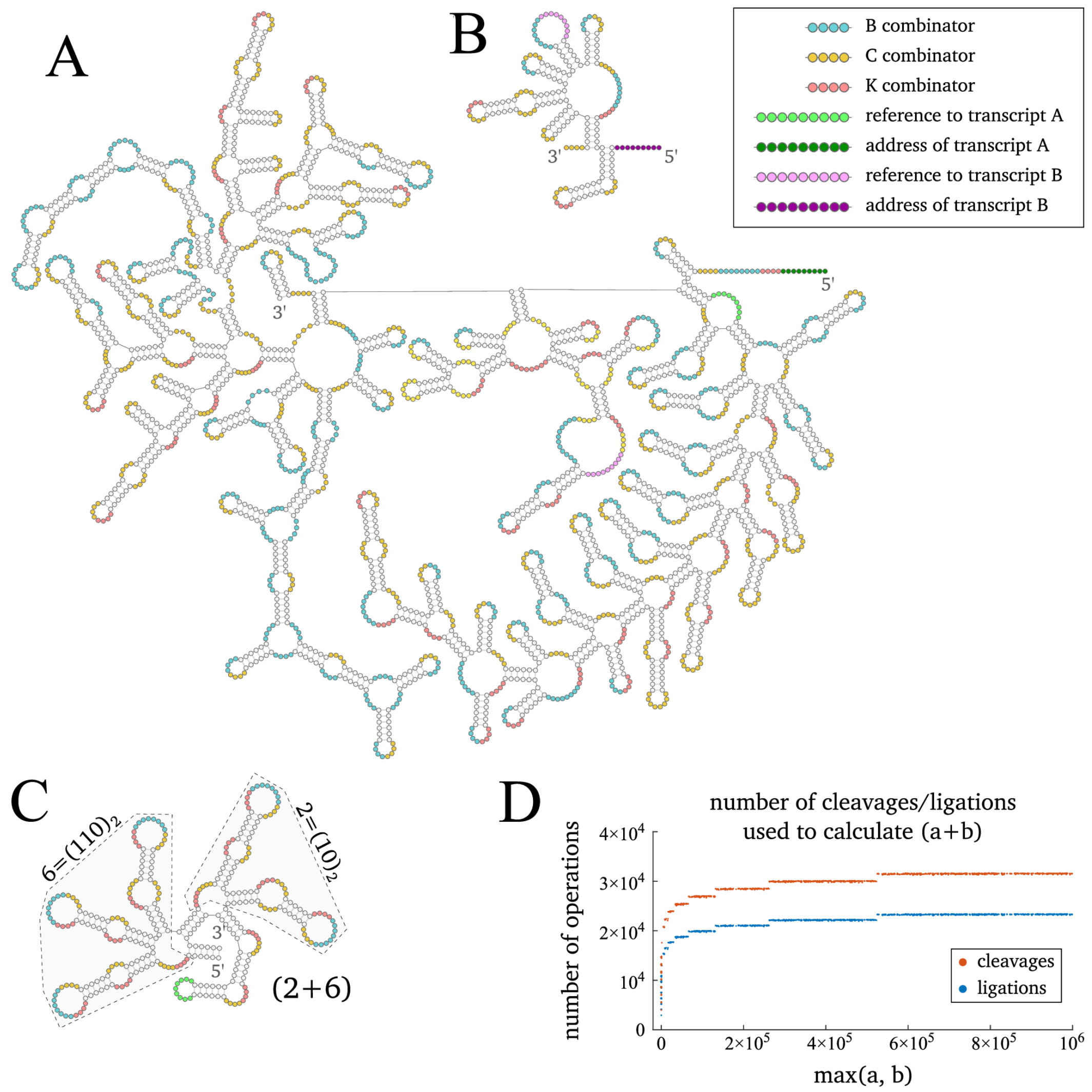

Figure 7: *Implementation of Addition* – An RNA program was constructed to add arbitrarily large numbers in logarithmic time using the editing rules of Model II (Fig 5) and an address/reference system similar that of Fig 6. This program involves two transcripts that are perpetually transcribed from static genomic templates and contain references to themselves. Since the **M** combinator is not used, the rules can be implemented by merely cleaving and ligating RNA strands at predefined positions. Combinator motifs were assumed to be 4-nt long and duplexes are each 6-bp. **(A)** Self-referencing 2,518-nt transcript responsible for addition that also contains a reference to transcript B. **(B)** Self-referencing 202-nt transcript responsible for reversing the content of a stack (or equivalently reversing the digits of a number). **(C)** an RNA program that computes 2+6 by referencing transcript A. **(D)** Simulation of the operation rules demonstrate that the number of operations increase logarithmically relative to the larger addend. (See Appendix C for derivations of the transcripts A and B).

To calculate 6+2=8, at least 4 copies of transcript A and 5 copies of transcript B need to be present. The final product is an RNA strand that contains four RNA elements corresponding to digits $(1000)_2$ which represents the number 8. The strength of this method becomes evident when examining how the number of operations scales with the size of the input. By simulating the editing rules for inputs up to $10^6$ I show that the number of operations scales logarithmically relative to the input (Fig 7D). (See supplementary data for simulation code).

The addition program in Figure 7 has arbitrary details and conventions (how to represent a number, which primitive combinators to use, etc). It was raised only for illustration purpose and is not intended to resemble how arithmetics is done in real biological cells. Our example shows a recursive program can be computed with purely cleavage, ligation, and transcription from static templates. A detailed proof (Appendix C section 7) shows that any computable function can be similarly computed using the same framework.

# 3. Discussion

This paper proposes a class of theoretical RNA-based models that are able to compute any computable function with identical algorithmic complexity to that of Turing machines. There is plenty of flexibility in the details of the models. The choice of primitive combinators is arbitrary and there are an infinite variety of primitives that lead to universal scope (e.g. SKI basis set or the BCKW basis set). There is also some freedom in other aspects of the model, such as the parenthesis convention (Figures 4 and 5 implement left-associative and right associative, respectively).

In one variation, I use an address/reference system to circumvent the requirement of duplicating RNA elements (which is needed for implementing the M or W combinators). Appendix C contains a proof that this model (which – strictly speaking – deviates from CL and λ-calculus because of its use of unbounded variables) is Turing-equivalent in scope and in algorithmic complexity (runtime and memory). A recursive function *f*, defined in terms of itself, can be executed by self-insertion of an RNA transcript representing *f*. In the process of computing *f* up to a depth of n, exactly n copies of the transcript are consumed and degraded. In a sense, the RNA-based RNA duplication of the purely CL models can be outsourced to the more plausible DNA-based RNA transcription. We already know the biological processes for perpetual RNA transcription from static genomic templates. In order to validate this model, what is left to discover are the specific enzymatic reactions that detect motifs and carry out the local cleavage/ligation operations at fixed positions relative to those motifs.

The components that are needed to implement these models are so familiar and unsophisticated that it appears plausible that nature may have already achieved an RNA-based universal computation system. In contrast to previous DNA/RNA-based models that achieve Turing-equivalence [42,56–59,64–72], these new models that bridge combinatory logic with RNA molecular biology are an attempt to conceive of a computation system within the limits of what may already exist in cells. This work motivates us to search for nature's universal computer at the molecular level and suggests some guiding details that may aid us in its discovery.

## 3.1. Challenges to the Theory

The formation of stem loops is an indispensable component of the models presented in this paper as it solves the otherwise difficult problem of parenthesis matching. But RNA secondary structure is not necessarily fixed or unique. The same RNA strand can change shape and fold into different kinetically stable structures [160]. An estimated 20-50% of mRNAs assume alternate conformations [161]. This can be a problem for a model that depends on stems representing parentheses because it makes RNA sequences ambiguous in their representations of CL terms. Even randomly generated nucleotide sequences lead to complex secondary structures that resemble some of the statistical properties of naturally occurring secondary structure, including non-local duplexes joining regions near the 3' and 5' ends [162,163]. Additionally, even if an RNA strand starts off in a unique highly stable state, local edits may radically change the thermodynamic energy landscape and lead to changes in secondary structure that do not reflect the intended operation rules of CL. Stable and controlled secondary structure is a critical part of the model. The argument for the theory's plausibility depends on stem loops representing nested terms.

How can the theory be reconciled with the unfettered nature of RNA structure? There are many factors that can influence the stability of secondary structure. Translation has a destabilizing effect on structure in mRNA [164]. RNA binding proteins or other factors may, in theory, stabilize stem loops as an RNA is being prepared to undergo CL modifications. Also, some sequences form more stable duplexes than others and it may be the case that only highly stable duplexes are treated as parentheses. G-C bonds are stronger than A-U bonds [165] and restrictions on the kinds of motifs that constitute parentheses can disambiguate them from the motifs that represent primitive combinators. And finally, there are many ways to construct the same CL term using primitive combinators. Semantically inconsequential variations to syntactic representation of terms can also be used to prevent the formation of unintended duplexes. For example, any number of identity combinators can be inserted after an open parenthesis without altering the intended computation, and that by itself may stabilize secondary structure. Importantly, none of the above models rely on strand displacement or the unwinding of already formed RNA duplexes. Once a RNA stem is formed, it never needs to come apart until that stem is excised and degraded. This makes RNA binding proteins an excellent candidate for stabilizing duplexes for the lifetime of an RNA strand. Conformational heterogeneity of RNA strands does pose a challenge to these models, but not an unsurmountable one.

Stochasticity may also occur at the level of primary structure. This can be dealt with by error-correction strategies. The most obvious error-correction method is programmatic error-correction, e.g. the use of check-sums or verification of program outputs. Additionally, an RNA based program can spawn many copies of the same sub-program in parallel. Some fraction of the programs may accumulate errors during computation but the overseeing program can select the most frequently occurring result. Combinatory logic is as powerful as any typical programming language and is capable of implementing any of the software-based error-correction strategies adopted in modern computing technology.

Another challenge for a solely RNA based molecular engram theory is RNA stability. If a molecule were to serve as a memory engram it must at least exhibit stability over similar time periods as cognitive memories. RNA molecules have an average half-life of around 7 hours [166,167]. DNA, however, can last for a lifetime and RNA information can be converted back to DNA through reverse transcription. In recent decades, we have learned that different cells of the same individual very often have differences in their DNA [168]. Neurons exhibit exceptionally high DNA diversity in what is known as neuronal *somatic mosacism* [169–172]. Various forms of DNA modifications exist in neurons, including single nucleotide variations, duplications, copy number variations, rearrangements, and aneuploidy. The quantitative extent of somatic mosaicism in the brain is currently under scrutiny [173,174] but it is especially pronounced in the human frontal cortex where average DNA content is enlarged by an estimated 4% [172] with a substantial fraction of neurons possessing unique massive copy number variations, on the scale of millions of basepairs long [170]. An important driver of somatic genomic diversity is *retrotransposition* [175], the process of transcription from DNA to RNA and subsequent *reverse transcription* from RNA back to DNA, resulting in duplications of specific genes. Until recently, it was not known if retrotransposition was restricted to mitotic progenitor cells, i.e., cells that divide to create more cells, or whether it also occurs naturally in post-mitotic cells [176,177]. This is important because reverse transcription is unlikely to facilitate cognitive memory storage unless it is widespread in healthy post-mitotic brain cells. Remarkably, somatic gene recombination, the process that has so far only been found in immune cells, has been recently reported in post-mitotic healthy human frontal cortex neurons [178–180]. The validity of this study is under debate and the results have not yet been independently replicated [181,182]. But an independent group has recently demonstrated retrotransposition in post-mitotic cultured neuronal cells [183]. These findings support the idea that RNA-based computation system can store the results of computation back into the form of DNA sequences as long-term memory. This idea had been previously formulated, leading to the prediction that *"individual neural cell will have distinctive spatially and temporally defined genomic sequences and chromatin structure"* [184]. In fact, most long non-coding RNA are localized in the nucleus and chromatin-associated [105,185–187]. Beyond neurons and lymphocytes, RNA-mediated genomic editing is known to occur in single-celled eukaryotes [188,189] and prokaryotes [190]. The long-term and short-term dichotomy of memory may be a reflection of two forms of RNA-based and DNA-based memory storage [184].

Can RNA modifications occur fast enough to potentially facilitate cognition? Two of the most well-studied RNA processes are transcription and translation. RNA Polymerase II transcribes RNA molecules at a rate of 18-100 nt/s equivalent to 36-200 bits/s [191–195]. And the ribosome translates RNA to protein at a speed of roughly 5-11 aa/s equivalent to 30-66 bits/s [195–199]. It is difficult to quantify how fast animals think but studies of different languages show that the information rate of human speech is roughly on the order of 40 bits/s (languages that are spoken faster have lower bits per syllable than languages that are spoken slower) [200]. These two information rates are within the same order of magnitude and also consistent with the spike rates of typical neurons (up to 100-200Hz). This means that RNA operations can in principle be fast enough to encode/transmit ideas communicated in speech as single RNA molecules. (See [201] for a discussion on the feasibility of recording spiking activity in the form of synthetic polynucleotides from an engineering perspective, rather than an evolutionary one). However, it remains to be shown that modifications of RNA molecules can occur fast enough to execute computation programs that underly animal cognition. In the models described in this paper, addition of two numbers requires on the order of $10^4$ cleavage/ligation operations. We do not know the speed at which these hypothetical editing rules may be implemented. But the spliceosome, which implements two cleavages and one ligation, takes an estimated 30 seconds to slice out an intron [202,203]. For an RNA-based computation system to facilitate cognition, either more efficient RNA programs must exist or RNA modifications must occur at extremely rapid rates.

## 3.2. Conclusion

I propose the theory that an RNA-based computation system exists in living cells that is Turing-equivalent, i.e., capable of computing any computable function. This is significant because life is confronted with problems of computation almost everywhere we look, from animal cognition, to single-cell decision-making and embryonic development. We have internalized the adequacy and the generality of universal computation in technology as we equip almost every device with a microprocessor. Likewise, an RNA-based universal computation system may be used to solve the many seemingly dissimilar problems of biology.

The specific models outlined in this paper are meant as a proof of principle that universal computation is well-within the reach of molecular biology without needing to invoke implausible molecular processes. The models implement combinatory logic through RNA molecules. Algorithmic complexities are identical to that of Turing machines (e.g., see Fig 1D). Parentheses are represented by reverse complementary sequences and RNA secondary structure elegantly solves the problem of parenthesis matching which is crucial to the operations of CL and λ-calculus. Universal computation is achieved through a decentralized process where distinct memory-less enzymes make local modifications to RNA molecules according to basic rules. The modifications can happen in parallel across many RNA strands – even at multiple locations within the same strand – and the order of operations do not matter. The system does not require a central processing unit or anything resembling a Turing machine tape head. No component of the model approaches the complexity of the ribosome. Instead, computation is carried out collectively by uncoordinated enzymes, which may themselves be catalytic or self-cleaving/self-ligating RNAs.

The fact that it is easy to imagine a universal computation system using RNA, while it is difficult to do so using network models, is highly suggestive that life's universal computation system – if there is one – resides in the subcellular domain and involves polynucleotides. An RNA-based computation system may have evolved in the very early stages of life – possibly before the evolution of DNA and proteins, consistent with the RNA-World theory [204,205]. (The third model depends on DNA-based transcription. But the first two models (Fig 4 & 5) do not necessarily depend on DNA or protein. Instead, they rely on RNA-dependent RNA replication, which is thought to have existed in the hypothetical RNA World). It is quite easy to stumble upon universal computation in systems that use a symbolic substrate that can grow in its number of symbols [30,31,33,34]. Once a universal computation system is established it is hard to see why it would be discarded throughout evolution. It is possibly the case that the RNA-based computation system is the very engine of evolution [206], optimizing mutations of offspring, consistent with recent evidence surrounding intergenerational inheritance of acquired traits [207,208].

Two major evolutionary events can be reinterpreted in the context of this theory. First, the evolution of complex multicellular organisms – which requires intracellular communication and sophisticated schemes of cooperation and division of labor [209–212] – potentially involved a general method of coordination across the RNA-based computation systems of different cells of the same somatic lineage. If the language of computation is encoded as RNA, it is reasonable to consider the theory that the messages conveyed across these cells are primarily in the form of RNA molecules. This is consistent with recent evidence on extracellular trafficking of RNA [213–217]. Second, the evolution of neurons and brains – which is thought to have occurred in independent lineages in metazoans [218–221] – may have served the purpose of rapid communication across cells, rather than serving as an entirely parallel computation system. Electrical/ionic signaling permits fast information transfer and coordinated motility in large multicellular organisms and has even evolved in plants for fast movement and decision-making based on information collected from sources that are many millimeters apart [222–225]. If prior to neurons the language of intracellular signaling was in the form of RNA, it is reasonable to consider the theory that neural signals encode RNA sequences that would have otherwise taken too long to export to downstream target cells. In this view, computation is primarily mediated through subcellular RNA-based processes, and the results of computations are then synaptically transmitted between cells to be used in other computations. This is consistent with the idea that the memory engram is in the form of polynucleotides – not synaptic plasticity – [35,44]. Of course, this is consistent with the existence of purely network based computations that do not directly involve RNA; network models can augment an RNA-based computation system.

The idea that polynucleotides are the substrate of memory can be traced back to the 1960s, when the discovery of DNA inspired a new scientific approach rooted in the hypothesis that the memory engram is a macromolecule [226–235]. It was hypothesized that "every idea is represented uniquely by a macromolecule with particular composition and sequence of monomer constituents" [235] and that "learning and memory depend on changes in genic material (or the by-products of genic activity) either in the nucleus or the cytoplasm of the nerve cell soma" [227], and that electrical signals in the nervous system are converted into nucleotide sequences through a hypothetical RNA transduction mechanism [236]. It was even suggested at the time that the molecular processes that underlie learning and encode new memories may be "*continuous across the phyla (as genetic codes are) and therefore would be reasonably similar for a protozoan and a mammal*" [2,237]. These ideas were largely abandoned in the 1970s [238–241] but have been rekindled in recent years [44,45,184,242–245]. The recent revival is rooted in the sobering realization that current theories of synaptic plasticity and network activity cannot explain learning, memory, and cognition [35,246] and that several lines of evidence bring into question the theory that synaptic strengthening/weakening is the primary form of long-term information storage in the brain [44,247–256].

While there are a number of challenges in applying this theory to cognitive memory, it is well-posed to resolve some of the challenges that current network models of cognition face in neuroscience. (See [45] for a comparison of the challenges of the connectionist/associative theories of memory and computational/representational theories of memory). One of the major criticisms of symbolic computation models has been that they are sequential, not parallel and distributed [257]. But the above models are unique in that while easily allowing recursion and nested structures, they are also highly parallel. A program can be split into numerous RNA strands, or even distributed along of the same RNA strand to be computed in parallel, thanks to the Church-Rosser theorem of λ-calculus and combinatory logic. More recently, combinatory logic has even been suggested as an alternative framework for understanding cognitive representations and learning [258].

The theory of natural RNA-based universal computation makes the falsifiable prediction that RNA molecules are modified in ways that radically deviate from their genomic templates and that these modifications are causally involved in cognition, cell-behavior, and/or development. Lengthening of RNA strands, through either duplication of RNA elements or integration of one strand into another, appears to be inevitable for fulfilling *memory expansion* (which is a necessary condition for any universal computation system in which memory usage is well-defined). We currently have very limited evidence for the existence of endogenous RNA sequences that cannot be mapped to the genome or accounting for through known RNA processing mechanisms. Some examples include widespread single-nucleotide variations (that cannot be attributed to ADAR or APOBEC enzymes) [259,260], non-genomically encoded 5'-poly(U) tails [155], exon repetition [153,154,156], post-transcriptional exon shuffling [132,133], and chimeric transcripts resulting from post-transcriptional fusion of RNA molecules [261–264] (although some of these results have been controversial [265–268]). The scarcity of evidence for extensive RNA editing is not definitive; detection of RNA sequences that are cell-specific or expressed in low numbers is notoriously difficult with current technology [110]. RNA sequence analysis pipelines typically discard reads that cannot be mapped to the genome or, in the case of *de novo* transcript assembly, discard aberrant reads that do not match other reads. It may be possible to discover the editing motifs of the hypothetical RNA-based system by careful examination of RNA sequencing data where unexpected editing patterns occur at low frequency. This is perhaps the simplest and most obvious first step for empirically testing the theory. To ultimately validate the theory, one needs to show that specific non-protein-coding RNA species are causally necessary for cognition and/or development, such that novel modifications to their sequences will have predictable effects in animal behavior and/or ontogeny. Conversely, the theory can be falsified if it is shown that the non-protein-coding content of the genome is mostly dispensable or can be replaced with random sequences without completely disrupting cognition, development, and cell function. But even if the RNA-based theory is disproved, the fact that universal computation is molecularly feasible and within the reach of evolution suggests that we must then search for life's universal computation system elsewhere.

## 4. Acknowledgements

I would like to thank Gaby Maimon (Rockefeller University), Jeremy Dittman (Weill Cornell Medicine), Abbas Rizvi (Columbia University), John Mattick (University of New South Wales), and Charles Randy Gallistel (Rutgers University) for the discussions and their insightful comments on this work.

# Appendix A: On the Theory of Computation

## 1. The Church-Turing Thesis and its Application to Biology

The Church-Turing thesis states that any effectively calculable function can be computed by a Turing machine. The thesis can be reformulated to replace "Turing machine" with any other equivalent computer (e.g. "lambda calculus term" or "primitive recursive function"). The thesis is not entirely well-defined and is subject to different interpretations because it relies on the informal notion of "effective calculability" (Copeland, 2020). Some consider the Turing machine to be the formalization of the notion of effective calculability, and therefore interpret the Church-Turing thesis as simply a definition (Soare, 1996). Others have taken the view that it is an empirical or mathematically provable statement that can in principle be proved or refuted. The Church-Turing thesis can be proved through an axiomatization of the notion of an "effectively calculable" (Dershowitz and Gurevich, 2008) (although one may argue over the specific set of axioms). I adopt the latter view in this paper and conclude from the Church-Turing thesis the following proposition: that *biological computation systems are limited by the Turing-equivalent scope of computation*. This proposition is empirical and can, in principle, be refuted. For instance, if one demonstrates that a neural network (that is itself physically realizable and describable in finite terms) can produce incomputable outputs from from computable inputs in such a way that can be harnessed for computation, that would disprove this proposition. (Note that this is not to say that any biological process can be simulated by a Turing machine. Rather, I only consider the proposition that any *system of computation* in biology, can be simulated by a Turing machine).

How can the scope of a biological computation system be evaluated and compared to that of Turing machines? The latter operates on strings of symbols whereas the former operates on physical material such as nerve impulses or molecular signals. Informally, computational scope can be compared by considering whether one computation system can solve the problems of another computation system. A formal framework of comparing computation systems having different domains has been developed that relies on the input and output domains being identical (Boker and Dershowitz, 2005, 2006, 2008). In other words, computation systems must be from strings to strings, or from numbers to numbers, or from nerve impulses to nerve impulses, etc. in order to be comparable. It may be possible to devise a more general framework that evaluates computation systems that have different input and output domains but we currently have no such framework. For the sake of argument, I will restrict the definition of biological computation system to only include systems that operate over the same input and output domain.

A high burden of justification must be met before we accept the idea that the abstractions appropriate for understanding computation in living organisms are exceptional and can somehow evade the structures and limitations that appear in the theory of computation.

# 2. The input/output/description domains must be effectively enumerable (computation over real numbers is pointless)

A computation system can be defined to take real numbers as inputs and output real numbers. Since classical computation systems (like the Turing machine) cannot operate on real numbers, they are incapable of simulating these systems to their full extent. This has been taken to mean that these real-number based systems are "super-Turing" or exceed the universal scope of computation. The problem with this argument is that it is relies on an aspect of the system that cannot effectively be harnessed and it is based on a lack of understanding about what the set of real numbers is.

Some sets are enumerable, meaning that each element can be referred to by a finite description like a string. The set of natural numbers is enumerable; each number can be shown by a string of digits. The set of rational number is also enumerable as each of its elements can be represented by a pair of numbers, both of which can be represented by strings of digits. The set of real numbers is not enumerable. That means that there are some real numbers that we cannot even refer to using human language or any particular notation system. The set of definable real numbers, is by definition enumerable. This set includes numbers like pi, the square root of two, and even some incomputable numbers like Chaitin's constant. But there are some particular numbers that we are unable to refer to, and no examples of such numbers can be given here since they are – by definition – impossible to refer to. By grasping this strange feature of real numbers, it becomes clear that it is impossible to harness the non-enumerability of the input/output domains of computation systems that operate over real numbers. It would not even be possible to report the outcome of a computational event involving undefinable real numbers, not through human language, and not through any formal notation system.

Furthermore, for a computation system to produce an undefinable real number as an output, it must have either received an undefinable input or have been based on an undefinable description because otherwise one can refer to the undefinable output in finite terms as follows: "the output produced by D when given input X". I believe it will be agreed that in any meaningful usage of a computation system the inputs and descriptions will be definable in finite terms. So the non-enumerable aspect of the description/input/output domains of computation cannot be harnessed.

Computation systems are allowed to consist of real-number variables or components. In fact having real valued components may even be an indispensable property of some computation systems. Here I am only concerned with the input/output/description domains and not the internal components of a computation system. It is even permissible to define a computation system that has real-numbers in its input/output/description, but such a computation system can be wrapped in another that is based on strings as inputs/outputs/description and converts the enumerable domains to the non-enumerable domains and vice versa, without effectively loosing any computation power.

[Some of the issues in this section have been previously discussed in (Davis, 2004, 2006a, 2006b)]

## 3. Continuous time and incomputable dynamical equations

It is commonly assumed that "effective computation" involves discrete steps in time. The axiomatization of algorithms by which it is possible to prove the Church-Turing thesis includes this discreteness of steps as an axiom (Dershowitz and Gurevich, 2008). By permitting continuous time, it may be possible to transcend Turing-computability. In fact the outcome of some dynamical systems cannot be predicted or simulated by Turing machines (Pour-El and Richards, 1979, 1981; Pour-El and Zhong, 1997; Smith, 2006). However, it is doubtful that such non-computable dynamical systems can be physically realized and harnessed in a computationally useful way (Weihrauch and Zhong, 1999, 2002). We currently have no conception of how a biological computation system could plausibly transcend Turing equivalence by taking advantage of continuous time. While remaining open-minded toward this possibility, I seriously doubt that it can ever be demonstrated to exist since any such demonstration would imply proof that the physical laws of the universe operate on continuous – rather than discrete - time.

## 4. The Finite/Infinite Classification of Machines is Misleading

It can be mistakenly thought that all physically relevant computation systems are weaker than Turing machines, on the grounds that Turing machines are infinite whereas physical systems are finite and the number of states they can assume are limited. It has even been claimed that because of this the appropriate model for physical systems is the finite state machine (van der Velde, 1993).

It is ironic that the finiteness of Turing machines needs to be justified considering that Turing was attempting to capture the notion of calculability by "finite means" (Turing, 1936) and that modern day computers are precisely the physical realization of universal computation inspired by Turing's formalization. Part of this misunderstanding may have stemmed from the title of Minksy's 1967 book "Finite and Infinite Machines". Minsky characterized Turing machines as infinite machines, while at the same time acknowledging that in order to "gain any really practical insight into real-life computers" one must study infinite machines.

But I do not find Misky's particular classification of machines into finite/infinite categories helpful. When we say Turing machines require unbounded memory we mean that *memory is unbounded by the system's description that specifies the input/output relation* not that it is unbounded by other things like the laws of physics.

When formalizing a system, one does not include all the real physical limitations in its abstract formalization. This is even true of the formalization of finite state machines. Finite state machines assume no bound on the size of the input or the amount of available time or free energy. If one imposes a predefined bound on input size, time to compute, or amount of consumable energy in the description of finite state machines, they will become as weak as a look-up table. Would it then be reasonable to say then that finite state machines are also physically unrealistic and irrelevant for studying real computation system? Physical limitations on time, energy, and space are issues that are dealt with independently from the problem of computation.

What allows Turing machines to be computationally superior to finite state machines is not that they begin with an infinitely long memory tape, but that memory is not bounded by the machine's description or operation rules.

A version of the Turing machine that begins with a finite sized memory tape but can add more memory squares when needed is just as powerful as the standard model of the Turing machine. The memory space of finite state machines, on the other hand, are limited by design and the bound is imposed by the description of the operation rules – not just by laws of physics. With this critical distinction in mind, it become clear that assuming unbounded memory is similar to assuming unbounded time or unbounded free energy. A physical implementation of a universal computer can – and will inevitably – have limitations on its memory capacity, energy consumption, and execution time but that does not mean it does not reflect the abstract model it is intended to implement.

# Appendix B: Computation Systems in Contemporary Biology

## 1. The Network Paradigm

In this section I characterize a model of computation which I believe serves as the most common framework in contemporary studies of natural computation. The network model, as defined below, is especially predominant in modeling cellular and neuronal computation. We use networks of genes that inhibit or promote one another to explain cell differentiation and embryonic development. We study networks of neurons or brain regions that inhibit or excite one another to produce models of cognition and animal behavior. And we attempt to understand cell behavior and embryonic development through studying the complex networks of genes, proteins, and enzymes that bidirectionally modulate one another's transcription rates, translation rates, decay rates, phosphorylation rates, and other quantities of cell chemistry.  Although it has never been explicitly stated, the network computation paradigm asserts that network models of computation are sufficient to achieve all forms of computation that are within the scope of living organisms.

Network models are typically schematized as a graph containing nodes and directed connections. Each node is thought of as containing a time-varying numeric value. Depending on what system is being used, a node's value can represent the activity of a brain region, the firing rate of a neuron, the membrane potential of a non-spiking neuron, the phosphorylation rate of a specific protein, the intracellular concentration of a molecule, or some other biologically relevant quantity. The connections of the graph correspond to the influences that those values have on one another. The network evolves over time in such a way that changes in any node's value is proportional to the sum of the values of its input nodes, usually including nonlinearities before and after the summation. The input to a model can be conceptualized as an input node that receives no inputs and simulates a predefined time-varying function. Likewise one or more node(s) in the network can be assigned as the *readout*, representing the result of the computation performed by the network. Sometimes their can be plasticity in the network. The weights of the connections (variables that control the strength of influence that one node has on another) can be modified according to some rule.

The network model can be understood as a dynamical system of finite dimensions. Each of the variables of the network (whether they are variables belonging to nodes or variables belonging to connections) can be assigned to one dimensions of the *state space*. At every point in time the network's state can be described by a single point in the finite dimensional state space. There is a function that determines how the system's state changes as time progresses. In other words, the rate of change of each of the variables can be written as a function of the network's overall state. Dynamical systems of finite dimensions are general enough to capture the network system. The critical assumption I make here is that the system is of finite dimensions. In other words, new nodes cannot be added during computation. Adding new connections between nodes is not of concern since one can assume a complete graph at the onset, with some of the connections being assigned zero weights to make them non-functional.

Dynamical systems have been studied as computational systems for several decades. There is a subtle difference in the dynamical systems that are used to study computation and the network model that is commonly used in biology. In dynamical systems, the input the to computation is encoded in the initial state of the system at time zero. To use those dynamical system to compute something, one needs to set the state to some value that represents the input, then evolve the system over time until it converges to a unchanging state and that state is pronounced as the output of the computation. Different starting points represent different inputs to the same computation. However, network models as computation systems are typically conceptualized as receiving a time varying input and producing a time varying output. Nevertheless I believe the studies and arguments regarding the limited scope of practically relevant dynamical systems are translatable to the network model. Let us now turn to the computational scope of dynamical systems that are finite dimensional and physically relevant.

## 2. The Computational Scope of Dynamical Systems

It is possible to simulate any Turing machine using dynamical systems (Asarin et al., 1995; Branicky, 1995; Graça et al., 2005; Koiran et al., 1994; Moore, 1990, 1991; Reif et al., 1990; Siegelmann and Sontag, 1991). The basic strategy for designing a dynamical system that simulates a Turing machine is to treat decimal places of a variable as a memory stack. So for example the number 0.31416... can be used as a stack containing the symbols '3', '1', '4', '1', '6', … on so on. The top of the stack can be read by multiplying by ten and rounding down. A symbol can be pushed into the stack by arithmetically adding a one digit number to the variable and then dividing the result by 10. And the stack can be popped by multiplying by 10 and subtracting a whole number. (In the cited examples, unary or binary representations were used instead of decimal representations). Using this clever method it is possible to simulate Turing machines with as few as three dimensions/variables with continuous time or two dimensions with discrete time steps (Moore, 1990, 1991), and even a single dimension at the cost of exponential slow-down of the simulation (Koiran and Moore, 1999). It is also possible to construct a neural network model that simulates any Turing machine, using three neurons that store memory stacks serving as the Turing machine memory tape and a network of neurons that act as the the Turing machine head by carrying out a predefined set of instruction rules (Siegelmann and Sontag, 1991). In all of these dynamical systems, the input to the computation is stored in the initial state of dynamical system at time zero. If a system is able to emulate any Turing machine, it is by definition capable of universal computation. So dynamical systems have universal computational scope.

However, all of these systems suffer from two practical weaknesses that make them irrelevant to understanding and implementing real physical systems. The first, regards the sensitivity of the system relative to small perturbations to its dynamical equations, i.e. the differential equations that govern how the system evolves over time. The second, regards the precision of the system's state variables, i.e. those variable that change with time. I will discuss each of these separately.

A dynamical system is *structurally stable* if it is surrounded by a neighborhood of homeomorphic systems. In other words, small additive changes to the function that govern how the system evolves over time will not affect the systems qualitative behavior (e.g. number of fixed points and number of periodic orbits). All of the cited systems that simulate Turing machines lack structural stability (Moore, 1998). (To intuitively understand why, consider the process of repeatedly pushing and popping a symbol '0' to the decimal memory stack variable. This involves repeatedly dividing by 10 and multiplying by 10. If the division factor and the multiplication factor differ by only a small amount, the memory variable will exponentially grow and never return to its original state. With a small error in the system's implementation resulting in a consistent multiplication factor of 10.01 and a consistent division factor of 9.99, after only ten repetitions of pushing and popping, the variable will increase in value by more than 2% and its memory content will become radically corrupted. For example 0.314159265... will turn into 0.320505705…)

The lack of structural stability means that *there is no arbitrarily small amount of perturbation to its differential equations that will preserve the system's general behavior*, even if the system has infinite precision and no noise once set into motion. This is almost like saying there is a neural network that carries out a desired computation but the synaptic weights have to be infinitely fine-tuned for it to work. Structural instability makes a model physically irrelevant, since even an implementation with an error of one part per billion would fail to resemble the intended general dynamics.

A conjecture by (Moore, 1998) states that *"No finite-dimensional system capable of universal computation is stable with respect to perturbations, or generic according to any reasonable definition"*. Moore's conjecture regards additive perturbations to the system's differential equations before it begins computing, not perturbations to the system's state as it is in the process of computing and not perturbations to the system's initial state. He uses the term "generic" to refer to systems of practical relevance. If Moore's conjecture is correct it follows that all practically relevant dynamical systems are computationally weaker than universal computers.

Let us now turn to the the second problem of practical concern. Any dynamical system with universal scope must assume unbounded precision. Many abstract models hinge on unboundedness, including the finite state machine model that assumes unbounded free energy or time to compute and the Turing machine model which in addition to unbounded time and energy assumes unbounded memory space. It can be argued that such systems are practically relevant so long as they can be implemented in such a way that physical limitations are never reached for all practical purposes. But this argument cannot be used for the memory space of dynamical systems. Extending the effective precision of a physical medium of information by one decimal point may require innovation and there is no algorithmic process that can iteratively achieve such a thing. Even if such a process were to be found it will very quickly hit Plank's constant within just a small number of iterations equivalent to a memory extension of roughly only a hundred bits ($2^{-110} \approx$ Planks constant in standard units). Whereas the memory of a Turing machine based physical system can easily be extended by recruiting more memory space on the hard disk of a computer. We have already extended the memory space of silicon computers by quadrillions of bits and we know it is possible to extend them by many more orders of magnitude before we hit any hard physical limits (Bekenstein, 1981). Likewise a computation system based on nucleotide polymers can expand its memory by simply adding nucleotide to the end of a polymer. Adding just five nucleotides would be equivalent to adding three decimal places to the effective precision of a dynamical system's variable, since $4^5 > 1000$.

The network model of computation is a computationally weaker system than universal computers. Its memory space is, for all practical matters, bounded as a function of is number of variables and it is unable to generate memory space during computation. Additionally, unless Moore's conjecture is proven wrong, no structurally stable dynamical system can simulate Turing machines, even if we assume infinite precision and zero noise. In agreement with (Moore, 1998) I believe that for a dynamical system to be practically relevant it must be structurally stable. So either living organisms have not achieved universal computation, or there is another model of computation that is more appropriate for understanding natural computation.

# Appendix C: λ-Calculus and Combinatory Logic Derivations

In this section I will introduce lambda calculus and combinatory logic, two very similar computation systems that are universal in scope. Both of them define a world of abstract entities that are essentially functions of functions to functions. There is a one-to-one mapping between the entities in lambda calculus (called lambda terms) and the entities in combinatory logic (called combinatory terms). The differences between these two systems are in their notations and the basic elementary operations used to evaluate terms. In lambda calculus the elementary operations are variable substitution and variable renaming. In combinatory logic the elementary operations are defined as a handful of functions called primitive combinators.

Since I find it easier to convey the power of functional programming with combinatory logic rather than λ-calculus, combinatory logic will be covered in more depth in this section and the examples of algorithms and data structures will all be given in terms of combinatory logic. Throughout this appendix, we will ultimately build a program capable of simulating any Turing machine. Importantly we will show that combinatory logic can simulate a Turing machine without any super-linear algorithmic slowdown. We will also show that this program, which is capable of computing any computable function, can avoid using any of the duplicating combinators (such as S, M, or W) as long as we can assign names to functions and reference functions by their names. Naming functions is not allowed in combinatory logic or λ-calculus. But this alternative system is useful for one of the RNA models in the main manuscript, which uses DNA-dependent RNA transcription instead of RNA-dependent RNA duplication.

## 1. Lambda Calculus

λ-calculus was invented by Alonzo Church as a formal system intended as a foundation for mathematics and logic (Church, 1932, 1933). It was his hope that a system based on the concept of functions instead of sets would make it possible to construct a complete formalism with no paradoxes and Gödel's incompleteness theorem would not apply to his formal system. His attempt failed – λ-calculus also contained inconsistencies that paralleled the paradoxes found in set theory (Kleene and Rosser, 1935) – but λ-calculus has emerged as an extremely powerful computation system with an undeniable impact on the fields of mathematics, logic, and the theory of computation. It has arguably inspired the creation of function programming languages and may perhaps deeply impact the field of molecular biology and provide the basis of understanding computation in nature.

In λ-calculus, everything is a function. Each function, referred to as a "λ-term" or "λ expression", can only take a single input and returns a single output. However, there is nothing for these λ expressions to operate on but other λ expressions. Simply put, everything is a function of functions to functions. It is an elegant feature of λ calculus that it does not start with any primitive functions. Functions do not have names and they are always referred by their complete λ-notations that fully describe what they do.

Without going into how λ expressions are built with λ-notations, let us construct a few example functions in English. The identity function is easy. We can create the identity function by defining it as "a function that returns whatever it is given" or "**a function that given any *x* returns *x***". Remember x is itself another function. So we can apply this function that we constructed to itself the output will be, unremarkably, itself. How about a constant function that gives a predefined output regardless of the input? Since we haven't created any other functions yet let us use the identity function as the constant. We define the following function: "a function that given any *x* returns the identity function". But remember that functions cannot have names. Instead of referring to the identity function by its name we must write its full description of what it does. Following this rule we can define our new function as "**a function that given any x returns the following function: a function that given any y returns y**". (Notice how we renamed the variable *x* in the identity function's definition to *y* in order to avoid variable name collision).

Even though lambda expressions only take one output, it is possible to construct effectively multivariate functions. To make a function that takes, say, three inputs we construct it so that it takes the first input and outputs a function that is supposed to take the next two inputs already storing the value of the first input in its definition. Inputs are then sequentially given to a λ-term. For example, the function that we most recently created can be interpreted as a function that takes two inputs, *x* and *y*, and returns the second one *y*. (This method is called currying, named after Haskell Curry). From this viewpoint, every function is actually a function that takes an infinite number of inputs because with every function it outputs another function that is ready for the next input.

Let us create another example that effectively takes in two inputs, but this time instead of ignoring the first and returning the second have it ignore the second and return the first. We can define it as follows: "**a function that given any *x* returns the following function: a function that given any *y* returns *x***". Even though naming functions is formally forbidden in λ-calculus, I will call this function the "kestral", and I will call the previous function, the function that ignores the first input and outputs the second, the "kite". So far we have made three functions, the identity function, the kestral, and the kite.

Now let us see what happens when we give the identity function as an input to the kestral. To do this we can replace "x" in the definition above with the identity function to get "a function that given any y returns the identity function". Again, functions cannot have names and "the identity function" is not allowed in the definition. We need to use its complete description ("a function that given x returns x") instead. The acceptable definition is "**a function that given any y returns the following function: a function that given any x return x**". Notice how the result is a function that we are already familiar with. It is a function that ignores the first input and returns the second input. If the kestral is given the identity function, it returns the kite function.

There are two very simple rules of operations in λ-calculus, called *α-conversion* and *β-reduction*. I will not go into the details of how they work as it would require explaining the λ-notations, but we already performed similar operations – variable renaming to avoid collision and variable substitution to compute the result of a function – in the examples above using an English notation. These two operations "reduce" the notation of an expression into another equivalent expression. Computation in λ-calculus can be understood as the iterative execution of these operations until a λ-expression cannot be further reduced (in technical terms we say the λ-expression is *"irreducible"*).

Lambda calculus was later shown to be equivalent to two other independently developed formal systems: general recursive functions and the Turing machine (Church, 1936; Kleene, 1936; Turing, 1937a, 1937b). From the basic rules above, one can create λ-terms that represent natural numbers (known as "Church numerals") and even construct λ-terms that add and multiply those terms. Despite its bare simplicity it has universal computational scope. The computational power will become more clear to the reader as we review examples algorithms, data structures, and arithmetic operations expressed in the language of combinatory logic.

## 2. Combinatory Logic (CL)

Combinatory logic (CL) preceded λ-calculus. It was invented by Moses Schönfinkel and later independently re-invented and further developed by Haskel Curry (Curry, 1930, 1932, 1934a, 1934b; Schönfinkel, 1924). CL is similar to λ-calculus with the main difference being that it starts with a set of primitive functions called "*basic combinators*". All functions are built from the primitive combinators. There is a one-to-one mapping between λ-terms and combinatory terms and the variety of functions that can be built are identical. The basic rules of reduction are slightly different, but these two systems are practically equivalent and CL is also a computation system with universal scope (i.e. Turing-equivalent).

Each defined combinator is represented by a single capital letter. We use lower case letters to denote variables that can be replaced by any combinatory term. The notation *xy* is equivalent to the return value of the function *x* when given *y*. So for example if we use the capital letter **I** to denote the identity function then we have ***I****x*=*x* for all *x*. Terms can be constructed from other terms using parentheses. The terms *x(yz)* and *(xy)z* are different in structure. In the first example, z is given as an input to y and the result is given as an input to x. In the second example, y is given to x and the resulting function receives z as an input. CL is conventionally left associative. In the absence of parentheses, the left-most two combinators must be evaluated first . So *wxyz* is equivalent to (((wx)y)z).

Multivariate functions can also be defined. The kestral combinator **K** is defined as **K**xy=x. But similar to λ-terms, CL terms must be conceptualized as functions that take only one input and return one output. For example, we defined **K** as though it takes two inputs, but the term **KI** is itself a valid combinatory term representing the "kite" function (see previous section for the equivalent constructions in λ-calculus). From the rules defined above, we have **KI**xy = (**KI**x)y = **I**y = y. So **KI** is a function that given any two inputs, ignores the first and outputs the second. **K** on the other hand is a function that given two inputs ignores the second and outputs the first. In general, any term can be given as an input to any term and there are no restrictions on how combinators can be combined.

There are many possible sets of basic combinators that can lead to a universal scope. We begin with two of the most commonly used sets: the BCKW system and SKI system[1]. The combinators **B**, **C**, **K**, and **W** are defined below.

```
Bxyz = x(yz)
Cxyz = xzy
Kxy = x
Wxy = xyy
```

The combinators S, K, and I are defined below.

```
Sxyz = xz(yz)
Kxy = x
```

1 The original basis set invented by Schönfinkel were B, C, I, K, and S. He recognized that the S and K combinators were sufficient to generate the other three. For more on the history of combinatory logic and lambda calculus see (Cardone and Hindley, 2006).

**I***x* = *x*

It can be shown that **S**, **K**, and **I**, can be expressed in terms of **B**, **C**, **K**, and **W**, and vice versa. For example, **B** can be replaced with **S(KS)K**. Let us prove this. We have to show that for any terms x, y, and z, S(KS)Kxyz is equal to Bxyz=x(yz). We do this by evaluating the term S(KS)Kxyz. A term is evaluated by iteratively applying the left-most combinators (*left-most* because of the omitted parentheses). We first apply the combinator S to the next three terms (KS), K, and x based on the rules above. S(KS)Kxyz=(KS)x(Kx)yz. We then drop the parentheses around (KS), because they are redundant, and evaluate the term KSx according to the rules above to obtain KSx=S. Continuing this method KSx(Kx)yz = S(Kx)yz = (Kx)z(yz) = Kxz(yz) = x(yz).

Either of these two systems are expressive enough to represent combinators in the other system, so any expression using combinators form one system is easily convertible to the other. For instance, every occurrence of B can be replaced with (S(KS)K). (In fact, the **I** combinator is redundant within the SKI system[2] and S and K are together a fully universal system). A computation program can be expressed as a combinator term built of the primitive combinators and parentheses. The input to a program *f* can either be a single term *x* (which itself can be composed of nested terms), or a sequence of *n* terms $x_1x_2x_3x_4...x_n$, depending on how the problem is framed. (It is possible to reframe a computation problem by composing the sequence $x_1x_2x_3x_4...x_n$ into a single term without destroying its content, and the corresponding solver *f* can be modified accordingly). Either way, the output is calculated by evaluating *fx* or $fx_1x_2x_3x_4...x_n$ , by iteratively applying the left-most term until it is no longer possible to do so. We shall now review examples on how to exploit this system for computation.

## 3. Booleans in Combinatory Logic

The following definitions can serve as a framework for carrying out circuit computations of boolean logic. There are not scalar values in CL. Everything is a combinatory term. We set the kestral **K** to correspond to true and the kite **KI** (or equivalently **CK**) to correspond to false. So if a boolean is followed by two values, the first value is chosen if that boolean is true and the second value if it is false. This method is known as the Church encoding of boolean values.

With trivial inspection it becomes clear that we already have an “if” construct. The algorithm “if condition x is true return y else return z” can simply be written as xyz, as long as x is a boolean in Church’s encoding. Adopting this method on can derive all unary and binary functions of boolean logic. Only the final solutions, not the derivations, appear below.

**NOT** (proper function): **C**(x)
**NOT** (postfix operand): x**(KI)K**
**OR** (proper function): **CIK**(x)(y)
**OR** (infix operand): x**K**(y)
**AND** (proper function): **CC(KI)**(x)(y)
**AND** (postfix operand): x(y)**(KI)**

2 This is easy to prove. For any *x* and any *y*, **SK***xy* = **K**y(xy) = y. Any two functions that give identical outputs for any input are equal by definition. From **SK**xy=y, we deduce **SK**x = **I** for any x. So **I** can be replaced by **SKS** or **SKK**.

```
                           AND (infix operand): xI(K(KI))(y)
                       NOR (proper function): C(CI(K(KI)))C(x)(y)
                           NOR (infix operand): x(K(KI))C(y)
                      NAND (proper function): C(CIC)(KK)(x)(y)
                      NAND (postfix operand): x(y)(KI)(KI)K
                          NAND (infix operand): xC(KK)(y)
                XOR (proper function operand): C(CIC)I(x)(y)
                           XOR (infix operand): xCI(y)
         XNOR / equality test (proper function): C(CII)C(x)(y)
           XNOR / equality test (infix operand): xIC(y)
   convert to standard form (postfix operand): xK(KI)
   convert to standard form (proper function): C(CIK)(KI)(x)
```

As an example let us compute the solution of `NOT((1 XOR 0 XOR 0 XOR 1) AND (1=0))`. In the expression below, NOT is encoded as a postfix operand, XOR as its multivariate form, equality (XNOR) as a proper function, and AND as a postfix operand.

```
K(C(CI(KI))K)I(KI)(C(CI(KI))K)I(KI)(C(CI(KI))K)IK(C(CII)(C(CI(KI))K)K(KI))(KI)(KI)K
```

To evaluate this function, it is sufficient to mindlessly apply the leftmost combinator at every step according to the rules of that combinator.

```
K(C(CI(KI))K)I(KI)(C(CI(KI))K)I(KI)(C(CI(KI))K)IK(C(CII)(C(CI(KI))K)K(KI))(KI)(KI)K
      = (C(CI(KI))K)(KI)(C(CI(KI))K)I(KI)(C(CI(KI))K)IK(C(CII)(C(CI(KI))K)K(KI))(KI)(KI)K
      = C(CI(KI))K(KI)(C(CI(KI))K)I(KI)(C(CI(KI))K)IK(C(CII)(C(CI(KI))K)K(KI))(KI)(KI)K
      = (CI(KI))(KI)K(C(CI(KI))K)I(KI)(C(CI(KI))K)IK(C(CII)(C(CI(KI))K)K(KI))(KI)(KI)K
      = CI(KI)(KI)K(C(CI(KI))K)I(KI)(C(CI(KI))K)IK(C(CII)(C(CI(KI))K)K(KI))(KI)(KI)K
      = I(KI)(KI)K(C(CI(KI))K)I(KI)(C(CI(KI))K)IK(C(CII)(C(CI(KI))K)K(KI))(KI)(KI)K
      = (KI)(KI)K(C(CI(KI))K)I(KI)(C(CI(KI))K)IK(C(CII)(C(CI(KI))K)K(KI))(KI)(KI)K
      = KI(KI)K(C(CI(KI))K)I(KI)(C(CI(KI))K)IK(C(CII)(C(CI(KI))K)K(KI))(KI)(KI)K
      = IK(C(CI(KI))K)I(KI)(C(CI(KI))K)IK(C(CII)(C(CI(KI))K)K(KI))(KI)(KI)K
      = K(C(CI(KI))K)I(KI)(C(CI(KI))K)IK(C(CII)(C(CI(KI))K)K(KI))(KI)(KI)K
      = (C(CI(KI))K)(KI)(C(CI(KI))K)IK(C(CII)(C(CI(KI))K)K(KI))(KI)(KI)K
      = C(CI(KI))K(KI)(C(CI(KI))K)IK(C(CII)(C(CI(KI))K)K(KI))(KI)(KI)K
      = (CI(KI))(KI)K(C(CI(KI))K)IK(C(CII)(C(CI(KI))K)K(KI))(KI)(KI)K
      = CI(KI)(KI)K(C(CI(KI))K)IK(C(CII)(C(CI(KI))K)K(KI))(KI)(KI)K
      = I(KI)(KI)K(C(CI(KI))K)IK(C(CII)(C(CI(KI))K)K(KI))(KI)(KI)K
      = (KI)(KI)K(C(CI(KI))K)IK(C(CII)(C(CI(KI))K)K(KI))(KI)(KI)K
      = KI(KI)K(C(CI(KI))K)IK(C(CII)(C(CI(KI))K)K(KI))(KI)(KI)K
      = IK(C(CI(KI))K)IK(C(CII)(C(CI(KI))K)K(KI))(KI)(KI)K
      = K(C(CI(KI))K)IK(C(CII)(C(CI(KI))K)K(KI))(KI)(KI)K
      = (C(CI(KI))K)K(C(CII)(C(CI(KI))K)K(KI))(KI)(KI)K
      = C(CI(KI))KK(C(CII)(C(CI(KI))K)K(KI))(KI)(KI)K
      = (CI(KI))KK(C(CII)(C(CI(KI))K)K(KI))(KI)(KI)K
      = CI(KI)KK(C(CII)(C(CI(KI))K)K(KI))(KI)(KI)K
      = IK(KI)K(C(CII)(C(CI(KI))K)K(KI))(KI)(KI)K
      = K(KI)K(C(CII)(C(CI(KI))K)K(KI))(KI)(KI)K
      = (KI)(C(CII)(C(CI(KI))K)K(KI))(KI)(KI)K
      = KI(C(CII)(C(CI(KI))K)K(KI))(KI)(KI)K
      = I(KI)(KI)K                                  * (equality check skipped here, automatically)
      = (KI)(KI)K
      = KI(KI)K
```

```
= IK
= K
```

The final result is **K**, equivalent to 'true'. It took 21 applications to compute, not counting the steps where leading parentheses are dropped. As soon as the result of the three XOR operations were calculated to be false it appropriately skipped calculating the result of the equality check, since the AND operation returns false if at least one of its arguments are false.

It is also possible to construct a proper function *f* that takes 6 unknown boolean inputs *u*, *v*, *w*, *x*, *y*, and *z* to compute **`NOT((u XOR v XOR w XOR x) AND (y=z))`**. The function *f* can be written as the expression below and used as *fuvwxyz* (although this is not necessarily the shortest possible expression of *f*).

```
B(B(B(B(B(B(C(B(BBB)(B(B(C(CI(KI))K))(B(CI(KI)))))(C(CII)(C(CI(KI))K))))(C(CI(C(CI(KI))K))I
)))(C(CI(C(CI(KI))K))I)))(C(CI(C(CI(KI))K))I)
```

What we have demonstrated so far is that *combinatorial logic* is possible to implement using *combinatory logic*. Despite their similar names, these two systems could not be further apart in terms of computation power. Combinatorial logic has a very narrow computational scope, weaker than finite state machines. Its abstract model has no states or memory and is not too different from a lookup table. CL, on the other hand, is as powerful as any computation system can be, capable of universal computation. Notice how it is possible to compute all logic gates with only a subset of the primitive combinators: B, C, K, and I[3]. Combinators S and W, the two primitive combinators that duplicate terms in their definition, have not yet been needed.

# 4. Data Structures

Church's encoding of pairs is arguably the simplest data structure that one can create in λ-calculus and CL . A pair is a data structure that stores two elements and allows the retrieval of either element. We will define Church pairs in terms of CL.

```
          constructor function (takes two arguments x and y): BC(CI)(x)(y)
       retrieve first element function (takes pair p as argument): CIK(p)
     retrieve first element postfix operand (operates on pair p): (p)K
    retrieve second element function (takes pair p as argument): CI(KI)(p)
  retrieve second element postfix operand (operates on pair p): (p)(KI)
```

Let us test the validity of this data structure by attempting to retrieve the second element of a pair. We assume `BC(CI)xy` as a pair and apply the *retrieve second element* function to it. At every step we simply apply the leftmost combinator. In order to make it easier to follow, leading parentheses are dropped explicitly in separate steps.

```
CI(KI)(BC(CI)xy) = I(BC(CI)xy)(KI) = (BC(CI)xy)(KI) = BC(CI)xy(KI) = C((CI)x)y(KI)
```

3 This is easy to prove formally by induction. Take the first input *x*. It is either true or false. The combinatory term *x(g)(h)* or equivalently *C(CI(g))(h)x* reduces to the function g if x is true and h if it is false. Both g an h can be constructed in terms of the subsequent inputs – independent of x. Any function of boolean combinatorial logic can be constructed as a composition of *if-then-else* statements. If x is the last input, g and h can be constructed as constants using K and (KI).

```
= ((CI)x)(KI)y = (CI)x(KI)y = CIx(KI)y = I(KI)xy = (KI)xy = KIxy = Iy = y
```

Note that *x* and *y* are variables and can be replaced with any term. There are no limits to the types of arguments that are given to the pair constructor function. Elements can even be pairs themselves. It is easy to see how one can store multiple terms using this design through nested pairs (e.g. pairs of pairs of pairs can store eight terms). Alternatively the definition for Church pairs can be generalized for k-tuples.

An issue with our design of pairs is that it is a single usage construct. The act of retrieving an element destroys the entire data structure. To retrieve multiple elements the pair needs to be duplicated using the **W** or **S** combinators. For example the function **W(B(CI(CIK))(CI(KI)))**, or equivalently **S(CI(KI))(CIK)**, takes a pair *p* of *x* and *y*, and applies the second element to the first element, *yx*[4].

Alternatively, a data structure can be designed that passes itself along after every function call. This can be advantageous if copying large terms is associated with high operational cost. We will next implement a stack data structure that embraces this strategy. Stacks have two essential functions: push and pop. The pop function works such that it takes two terms *f* and *e* and a stack *s*. If the stack is empty it returns *e*. Otherwise it returns $fxs^{-1}$ where *x* is the element at the top of the stack and $s^{-1}$ is the stack with the top element removed[5]. In order to simply discard the top element, the pop function can be called with *f*=**KI** and e=**KI**. To check whether the stack is empty on can call the pop function such that it includes a push function. We construct an *empty check* function that takes two terms *f* and *e* and stack *s*. If the stack is empty it returns e. Otherwise it returns *fs*.

| | |
|---:|:---|
| **empty stack**: | **KI** |
| **pop** (takes functions f & e and stack s): | **BC(CI)**(f)(e)(s) |
| **push** (takes element x and stack s): | **BC(C(B(BK)))**(x)(s) |
| **pop and discard element** (takes stack s): | **BC(CI)(KI)(KI)**(s) |
| **empty check** (takes functions f & e and stack s): | **C(B(B(BC(CI)))B)(BC(C(B(BK))))**(f)(e)(s) |

For instance, pushing three elements *x*, *y*, and *z* into an empty stack results in the following term.

```
C(C(B(BK))z)(C(C(B(BK))y)(C(C(B(BK))x)(KI)))
```

We can then observe what happens when we call the pop and discard element function on this stack. The pop and discard element function is emphasized in bold. The remainder of the stack is underlined to show that it remains unperturbed during this operation. After 11 applications, not including steps where redundant parentheses are removed, the result is a stack that contains elements *y* and *x*.

```
BC(CI)(KI)(KI)(C(C(B(BK))z)(C(C(B(BK))y)(C(C(B(BK))x)(KI))))
       = C((CI)(KI))(KI)(C(C(B(BK))z)(C(C(B(BK))y)(C(C(B(BK))x)(KI))))
       = ((CI)(KI))(C(C(B(BK))z)(C(C(B(BK))y)(C(C(B(BK))x)(KI)))))(KI)
       = (CI)(KI)(C(C(B(BK))z)(C(C(B(BK))y)(C(C(B(BK))x)(KI)))))(KI)
       = CI(KI)(C(C(B(BK))z)(C(C(B(BK))y)(C(C(B(BK))x)(KI)))))(KI)
       = I(C(C(B(BK))z)(C(C(B(BK))y)(C(C(B(BK))x)(KI)))))(KI)(KI)
```

4 It is possible to accomplish this while avoiding duplication. The **CI(CI)** term takes a pair and applies the second element to the first. But this function bypasses the interface functions that we defined and the general point still remains that the data structure with all its elements is destroyed every time a function is applied to it.

5 Here we are employing a strategy inspired by the Mogensen-Scott encoding method. That is, the stack has two states: either empty or non-empty. The functions corresponding to each state is given to the stack separately.

```
= (C(C(B(BK))z)(C(C(B(BK))y)(C(C(B(BK))x)(KI))))(KI)(KI)
= C(C(B(BK))z)(C(C(B(BK))y)(C(C(B(BK))x)(KI)))(KI)(KI)
= (C(B(BK))z)(KI)(C(C(B(BK))y)(C(C(B(BK))x)(KI)))(KI)
= C(B(BK))z(KI)(C(C(B(BK))y)(C(C(B(BK))x)(KI)))(KI)
= (B(BK))(KI)z(C(C(B(BK))y)(C(C(B(BK))x)(KI)))(KI)
= B(BK)(KI)z(C(C(B(BK))y)(C(C(B(BK))x)(KI)))(KI)
= (BK)((KI)z)(C(C(B(BK))y)(C(C(B(BK))x)(KI)))(KI)
= BK((KI)z)(C(C(B(BK))y)(C(C(B(BK))x)(KI)))(KI)
= K(((KI)z)(C(C(B(BK))y)(C(C(B(BK))x)(KI))))(KI)
= (((KI)z)(C(C(B(BK))y)(C(C(B(BK))x)(KI))))
= ((KI)z)(C(C(B(BK))y)(C(C(B(BK))x)(KI)))
= (KI)z(C(C(B(BK))y)(C(C(B(BK))x)(KI)))
= KIz(C(C(B(BK))y)(C(C(B(BK))x)(KI)))
= I(C(C(B(BK))y)(C(C(B(BK))x)(KI)))
= (C(C(B(BK))y)(C(C(B(BK))x)(KI)))
= C(C(B(BK))y)(C(C(B(BK))x)(KI))
```

No copying combinator (neither **W** nor **S**) is involved in these functions and the deeper elements in the stack are untouched and fully preserved with every call. We can also write a "top element" function that works similar to the pop function but effectively leaves the stack unchanged. This is implemented by combining the pop function with a function that duplicates the top element, pushing one copy back into the stack and feeding the other copy to function *f*. The *top* function includes a single duplicating combinator term (**W** in one implementation and **S** in the other) but that term is only applied to the popped element. The entire stack is not duplicated, only the top element is[6].

```
top (takes f and e and stack s): B(B(B(B(BC(CI))W)(B(CI(BC(C(B(BK)))))))(BB))(BB) f e s
                             or: B(B(C(B(BC(CI)))(BC(C(B(BK)))))S)(BB) f e s
```

## 5. Recursion

CL is capable of implementing recursive functions. We demonstrate this in the examples below, using the stack that was designed in the previous section.

Let us first solve the following problem: "Given a stack consisting of Church booleans true=**K** and false=**KI**, remove all the consecutive *true* elemoents from the top of the stack". We take a recursive approach and seek an algorithm *q* that implements: "if the stack has a top element equal to *true* then pop that value and execute algorithm *q* on the remaining stack - otherwise return the stack". To investigate the value of the top element it must first be *popped* out of the stack. So the first step of algorithm *q* is the pop function. In combinatory terms:

```
q = BC(CI)fe
```

We did not include s in the definition because *q* must itself take a stack as an input (*rs* = **BC(CI)***fes*). If the stack is empty, we want the algorithm to return an empty stack. So we must set *e* to be equal to an empty stack **KI**. But what must *f* be set to? If the stack is not empty, *q* will reduce to the term $fxs^{-1}$, where *x* is the top value and $s^{-1}$ is the stack with the stack popped once. The function f must push *x* back into $s^{-1}$ if *x* is false, and it should execute algorithm *q* on $s^{-1}$ if *x* is true. If we set *f*=`BC(CI)`$f_T f_F$, it will execute $f_T$ if $f_F$ is true and execute $f_F$ if *x* is false (since

6 The duplicating combinator can be avoided if we know that the stack consists of Church booleans **K** and **KI**. For boolean inputs the W combinator can be replaced with **C(C(CI(C(CIK)K))(C(CI(KI))(KI)))**.

BC(CI)$f_T f_F x = x f_T f_F$). The functoin $f_T$ can simply be set to $q$ and $f_F$ can be set to **BC(C(B(BK)))(KI)** which is the push function with **(KI)** as an argument (pushes false into the stack). Substituting the solved variables we have: $f$=**BC(CI)**$q$**(BC(C(B(BK)))(KI))**. So $q$ can be rewritten as:

```
q = BC(CI)(BC(CI)q(BC(C(B(BK)))(KI)))(KI)
```

But this is not a valid definition for $q$ because it is defined in terms of itself. To solve this recursive equation we turn to the **Y** combinator. The **Y** combinator takes any function $g$ and solves the equation $x=gx$ for $x$. In other words the term **Y**$g$ fulfills the recursive equation such that **Y**$g$=$g$(**Y**$g$). The **Y** combinator can be expressed in terms of our primitive combinators: **Y**=**B(WI)(BWB)**. To solve for $q$, we first rewrite it in the form $q=gq$. This is always possible to do for any function defined in terms of itself.

```
q = BC(CI)(BC(CI)q(BC(C(B(BK)))(KI)))(KI)
  = C(BC(CI))(KI)(BC(CI)q(BC(C(B(BK)))(KI)))
  = B(C(BC(CI))(KI))(BC(CI)q)(BC(C(B(BK)))(KI))
  = C(B(C(BC(CI))(KI)))(BC(C(B(BK)))(KI))(BC(CI)q)
  = B(C(B(C(BC(CI))(KI)))(BC(C(B(BK)))(KI)))(BC(CI)) q
  = Y (B(C(B(C(BC(CI))(KI)))(BC(C(B(BK)))(KI)))(BC(CI)))
  = B(WI)(BWB)((B(C(B(C(BC(CI))(KI)))(BC(C(B(BK)))(KI)))(BC(CI))))
```

This is the final solution to the problem. An example application of this function is too lengthy to be included in this paper. The use of the **W** combinator that duplicates terms is unavoidable in solving recursive problems. The explanation for why the **Y** combinator works the way it does falls outside of the scope of this paper. We suffice to show how it can be used to solve recursive equations of CL.

We will go through one more example, this time dealing with more than one data structure simultaneously. The problem we shall solve is to reverse the content of a given stack. This can be done by creating an empty stack and popping the elements of the original stack and pushing them into the new stack iteratively, until the original stack is empty. This time we define our algorithm as $r$(KI), such that $r$ is a yet unconstructed term that takes two stacks and empties all the content of the second stack into the first stack. We already provide the first stack as an empty stack (KI) and declare $r$(KI) as the algorithm that solves the stack reversal problem. Now what must r do at each step? First, what is the non-recursive part of the algorithm?

Given two stacks $s_1$ and $s_2$, $r$ must first swap them (to make $s_2$ accessible for the next operation) and then call the pop function on $s_2$. We define this in terms of unknown function $f$ and $e$:

$$rs_1s_2 = C(\mathbf{BC(CI)}fe)s_1s_2 = \mathbf{BC(CI)}fes_2s_1$$

If the stack $s_2$ is empty, the above term will reduce to $es_1$. We can set $e$=I since the algorithm should terminate and return $s_1$ when $s_2$ is empty. If $s_2$ is not empty, the above term reduces to $fxs_2^{-1}s_1$ where $s_2^{-1}$ is the stack $s_2$ after one pop operation and $x$ is the top element of $s_2$. At this point $f$ must swap the two stacks (to make $s_1$ accessible for the next operation) and push $x$ into $s_1$ and then execute $r$ on the resulting two stacks. So we can define $f$=BC(B(Br)<push>)=BC(B(Br)(BC(C(B(BK))))). We validate that $f$ works as intended . For readability the push function BC(C(B(BK))) is abbreviated as <push>.

$$fxs_2^{-1}s_1 = BC(B(Br)<push>)xs_2^{-1}s_1 = C(B(Br)<push>x)s_2^{-1}s_1 = B(Br)<push>xs_1s_2^{-1} = Br(<push>x)s_1s_2^{-1}$$
$$= r(<push>xs_1)s_2^{-1}$$

Substituting $f$ and $e$ in the definition of $r$ and solving for r using the **Y** combinator we get:

```
r = C(BC(CI)fe) = C(BC(CI)fI)
  = C(BC(CI)(BC(B(Br)(BC(C(B(BK))))))I)
  =   B(B(C(B(B(B(C(BC)I)(BC(CI)))(BC)))(BC(C(B(BK)))))B)B r
  = Y(B(B(C(B(B(B(C(BC)I)(BC(CI)))(BC)))(BC(C(B(BK)))))B)B)
  = B(WI)(BWB)(B(B(C(B(B(B(C(BC)I)(BC(CI)))(BC)))(BC(C(B(BK)))))B)B)
```

So the combinatory term that solves the stack reversal problem is:

```
r(KI) = B(WI)(BWB)(B(B(C(B(B(B(C(BC)I)(BC(CI)))(BC)))(BC(C(B(BK)))))B)B)(KI)
```

Readers can run the above program on a stack of their choice. Carrying out the operations by hand may be tedious, so using a CL interpreter is encouraged. This program reverses a stack of three elements in 98 applications.

## 6. Numbers and Arithmetic Operations

A common numeric system in λ-calculus/CL is Church numerals. According to this system a number $n$ is a combinator term $N_n$ that takes function $f$ and term $x$ and applies $f$ to $x$, $n$ times. For example $N_1 fx = fx$, $N_3 fx = f(f(fx))$, and $N_0 fx = x$. In terms of the primitive combinators that we defined earlier, $N_0$=**KI**, $N_1$=**I**, $N_2$=**SBI**, $N_3$=**SB(SBI)**, $N_4$=**SB(SB(SBI))**, and so on. The function **SB** increments a number by 1. The function **BS(BB)** takes two numbers and returns the sum, multiplication is simply the term **B**, and exponentiation is simply the term **I**. Similarly, there are functions for subtraction, division, comparison, and equality checks. But Church's numeric system is a unary system where the length of numerical representation grows linearly with the number. It is undoubtedly useful for studying number theory but if the number of operations and memory space is of concern, arithmetic operations are inefficient.

The prevalence of unary encoding systems has given λ-calculus/CL a reputation for being slow. But numerals can be encoded in binary or any base of choice and arithmetic operations can be implemented at optimal complexity (Mogensen, 2001).

It may already be clear to the reader that the stack data structure introduced above can be used to store binary digits of a number with the least significant digit at the top of the stack. Addition can be implemented as a recursive algorithm that takes two numbers (two stacks), pops the least significant digits, adds them, and pushes them into a new stack, carrying the overflow digit over to the next addition operation and reversing the result stack when the two input stacks are empty. Multiplication by 2 can be done by pushing a zero digit to the stack. Multiplication of two arbitrary numbers can be implemented by iteratively popping the digits of one stack and pushing zeros to the second stack and numerically adding the entire number of the second stack to a result stack if the first stack yielded a 1 digit with the pop operation. With numerals as stacks, there are no limits to the number of digits in a number. But it is also possible to encode numbers as having a fixed number of digits, similar to the standard 32 bit encoding for integers in computers. Boolean logic circuits can be designed for addition, multiplication, subtraction, and division for fixed size numbers, in the same way modern day CPUs implement arithmetic operations.

## 6.1. Derivation of an Addition Program

Below is the derivation for a program that implements addition by taking two stacks of booleans that represent the binary expansion of two numbers. We derive $\sigma$ such that $\sigma\omega s_3 s_1 s_2$ is equal to $r\langle push\rangle\omega s_3$ if the two stacks $s_1$ and $s_2$ are empty. Otherwise it pops $s_1$ and $s_2$ and adds their top elements with the carry $\omega$, pushes a digit into $s_3$ and stores the new carry as a new $\omega$, returning the new $\sigma\omega s_3 s_1 s_2$.

```
<ADDITION> := σ(CK)(CK)

<ADDITION>(s_1)(s_2) = σ(CK)(CK)(s_1)(s_2)

σ := (π_0)(g)(π_1)(g)(π_2)(h_0)

π_0 :=
B(B(B(B(B(B(B(B(B(B(BC)C))C))C))C))C)(B(B(B(B(B(B(BC)C))))))

π_1 :=
B(B(B(B(B(B(B(B(B(BC)C))B)B))C))C)(C(B(B(BC(B(BC(BC)))))))

π_2 :=
B(BC(B(B(BC)C)))(C(B(B(B(BC(B(B(B(BC(B(B(B(BC(B(BCB)))C)B)))C)B)))C)B)(C(BC(B(BC(BC)))))))

π_0 x_1 x_2 x_3 x_4 x_5 x_6 x_7 x_8 x_9 = x_1 x_8 x_2 x_3 x_4 x_5 x_9 x_6 x_7
π_1 x_1 x_2 x_3 x_4 x_5 x_6 x_7 x_8 x_9 = x_4 x_7 x_5 x_1 x_2 x_6 x_8 x_9 x_3
π_2 x_1 x_2 x_3 x_4 x_5 x_6 x_7 x_8 x_9 = x_6 x_1 x_4 x_2 x_5 x_7 x_8 x_9 x_3

(π_0)(g)(π_1)(g)(π_2)(h_0)(ω)(s_3)(s_1)(s_2) = (g)(s_1)(π_1)(g)(π_2)(h_0)(s_2)(ω)(s_3)

g := <pop>(f)(e)

<pop> := BC(CI)
f := B(B(BC)C)(CI)(CK)
e := B(B(BC)C)(CI)K(CK)(CK)

fx_1 x_2 x_3 = x_3(CK)x_1 x_2        [in this case ∅_i=false and t_i=x_1]
ex_1 x_2 = x_2 K(CK)x_1              [in this case ∅_i=true and t_i=false]

g(s_i)(π_j) = (π_j)(∅_i)(t_i)(s_i^-1)

(g)(s_1)(π_1)(g)(π_2)(h_0)(s_2)(ω)(s_3)
= (π_1)(∅_1)(t_1)(s_1^-1)(g)(π_2)(h_0)(s_2)(ω)(s_3)
= (g)(s_2)(π_2)(∅_1)(t_1)(h_0)(ω)(s_3)(s_1^-1)
= (π_2)(∅_2)(t_2)(s_2^-1)(∅_1)(t_1)(h_0)(ω)(s_3)(s_1^-1)
= (h_0)(∅_2)(∅_1)(t_2)(t_1)(ω)(s_3)(s_1^-1)(s_2^-1)

h_0 := C(BC(C(BC(CC(CK)))(h_1)))(h_2)

(h_0)(∅_i)(∅_j) = (∅_i)(∅_j)K(h_1)(h_2)          [if (∅_i & ∅_j) then h_1 otherwise h_2]

h_1 := K(K(C(C(B(B(B(B((BK)K)(r(CK))))))(<push>K))I))

<push> := BC(C(B(BK)))

r := C(BC(CI)(BC(B(Br)(BC(C(B(BK))))))I)
   = B(WI)(BWB)(B(B(C(B(B(B(C(BC)I)(BC(CI)))(BC)))(BC(C(B(BK)))))B)B)

                                   [r(CK)(s) returns stack s reversed]
```

$$h_1(t_2)(t_1)(\omega)(s_3)(s_1^{-1})(s_2^{-1}) = r(CK)(\omega(<push>K)(I)(s_3))$$

$$h_2(t_2)(t_1)(\omega) := (t_2)(a_1)(a_0)(t_1)(b_2)(b_1)(b_0)(\omega)(c_3)(c_2)(c_1)(c_0)(\pi_3)(\sigma)(<push>)$$

$$\begin{aligned}
a_0 &:= (B(BK)) && \textit{[sum is 0 so far]}\\
a_1 &:= K && \textit{[sum is 1 so far]}\\
b_2 &:= (B(B(BKK))) && \textit{[sum is 2 so far]}\\
b_1 &:= (BK(B(BK))) && \textit{[sum is 1 so far]}\\
b_0 &:= (BKK) && \textit{[sum is 0 so far]}\\
c_3 &:= (BC(CC)KK) && \textit{[sum is 3]}\\
c_2 &:= (BC(CC)K(CK)) && \textit{[sum is 2]}\\
c_1 &:= (BC(CC)(CK)K) && \textit{[sum is 1]}\\
c_0 &:= (BC(CC)(CK)(CK)) && \textit{[sum is 0]}
\end{aligned}$$

$$\pi_3 x_1 x_2 x_3 x_4 x_5 := x_3 x_1 (x_4 x_1 x_5)$$

$$h2(t_2)(t_1)(\omega)(s_3)(s_1^{-1})(s_2^{-1}) = \sigma(\omega_{new})(s_3^{+(t1 \text{ xor } t2 \text{ xor } \omega)})(s_1^{-1})(s_2^{-1})$$

$$\textit{[}\omega_{new} = (t_1+t_2+\omega > 1)\textit{]}$$

```
=> σ =
BC(C(C(C(C(C(B(B(B(B(B(B(BC)C)))))(C(CI(B(B(BC)C)(CI)(CK)))(B(B(BC)C)(CI)K(CK)(CK))))
(B(B(B(B(B(B(B(B(BC)C))B)B))C))C)(C(B(B(BC(B(BC(BC)))))))))(C(CI(B(B(BC)C)(CI)(CK)))
(B(B(BC)C)(CI)K(CK)(CK))))(B(BC(B(B(BC)C)))
(C(B(B(B(BC(B(B(B(BC(B(B(B(BC(B(BCB)))C)B)))C)B)))C)B)(C(BC(B(BC(BC))))))))
(C(BC(C(BC(CC(CK)))(K(K(C(C(B(B(B(B(BKK)(r(CK))))))(C(C(B(BK))K)))I)))))(C(B(BC(BC))
(C(B(BC(BC))(C(B(BC(BC))(C(B(BC(BC))(C(B(BC(BC))(C(B(BC(BC))(C(B(BC(BC))
(C(BC(C(BC(C(BC(C(CI(B(BK)))K))(B(B(BKK)))))(BK(B(BK)))))(BKK)))(C(CCK)K)))(C(CCK)(CK))))
(C(CC(CK))K)))(C(CC(CK))(CK))))(B(BCB)(C(B(BCB)B)))))σ))(BC(C(B(BK)))))))
```

In the above representation σ is defined in terms of itself. Like the previous examples, the self-referencing can be eliminated by using the Y combinator. For the purposes of the manuscript we will not derive the proper full form of σ. Instead we will show what it looks like if we switch from left-associative combinator logic to right-associative CL using brackets instead of parentheses. The expression below is equivalent to the RNA representation of Fig 7A in the manuscript.

```
σ =
[BC]C[C[C[C[C[BBBBB[BBC]C][C[C[CK]C][[B[BBC]C]C[CK]C]CK][[C][[C][C[CK]C]K]CK]CK]
[B[BB[BB[B[BB[BBC]C]B]B]C]C]CBB[BC]B[BC]BC][C[C[CK]C][[B[BBC]C]C[CK]C]CK][[C][[C]
[C[CK]C]K]CK]CK][B[BC]B[BBC]C]C[B[B[B[BC]B[B[B[BC]B[B[B[BC]B[BC]B]C]B]C]B]C]B]C[BC]B[BC]BC]
[C[BC][C[BC][CC]CK]KK[C[CBBB[B[BK]K]rCK]C[CBBK]K][CK]C][C[B[BC]BC][C[B[BC]BC][C[B[BC]BC]
[C[B[BC]BC][C[B[BC]BC][C[B[BC]BC][C[B[BC]BC][C[BC][C[BC][C[BC][C[C[CK]C]BBK]K]BB[BK]K]
[BK]BBK][BK]K][C[CC]K]K][C[CC]K]CK][C[CC]CK]K][C[CC]CK]CK][B[BC]B]C[B[BC]B]B]σ][BC]CBBK
```

```
r = C[[C][C[CK]C][BC][BBr][BC]CBBK][[CK]C]
```

## 7. Universal Computation

It is possible to simulate any Turing machine using CL, with linearly proportional runtime. Using the general guidelines above, one can assume two stacks of binary values that serve as a memory tape. The tape head can be conceptualized as a finite state machine. A recursive program can be written that at each step pops the first stack, decides what do do with it, permutes the list of states accordingly, pushes a value back into the stack and if needed, and swaps the two stacks as needed. Let us construct this program step by step.

The program $\varphi$ simulates a Turing machine with $N$ states (not including the halt state) and an alphabet of $M$ characters (including the blank character). The program takes three inputs: the current state of the tape head, followed by the left and right tape contents as two stacks.

$$\varphi c s_1 s_2$$

To encode the state as $c$, we will use a multiplexer function designed to take $N$ inputs and select the $i$-th input. The multiplexer is a function that will come in handy, so let us formalize and construct it.

$$m_{r,n} x_1 x_2 \ldots x_r \ldots x_n = x_r$$

The construction of our multiplexer will be by induction. We use the following base cases and induction steps.

$$m_{1,1} := I$$
$$m_{1,n} := BK(m_{1,n-1})$$
$$m_{r,n} := K(m_{r-1,n-1})$$

For example $m_{2,2} = K(m_{1,1}) = KI$, or $m_{2,3} = K(m_{1,2}) = K(BK(m_{1,1})) = K(BKI)$. (For simplicity, one can include an additional base case $m_{1,2} = K$).

The state $c$ can be set as $m_{i,N}$. The current state of the tape head is the $i$-th state of the $N$ possible states.

Additionally the stacks contain one multiplexer function for each of the characters written on the tape. The $k$-th character from the alphabet of $M$ characters is represented by $m_{k,M}$. We can assume that the blank character is $m_{1,M}$.

We now can begin the formal construction of the Turing machine simulator program. We first define $\varphi$ in terms of itself.

$$\varphi := \pi_1 f_1 f_2 \ldots f_N \varphi$$

The functions $f1$ to $fN$ will be defined later and correspond to each of the $N$ states of the finite state machine of the tape head. $\pi_1$ is a permuting combinator that pulls $c$ to the front and send $\varphi$ to the back. We will not construct $\pi_1$ but formally:

$$\pi_1 x_1 x_2 \ldots x_N x_{N+1} x_{N+2} x_{N+3} x_{N+4} = x_{N+2} x_1 x_2 \ldots x_N x_{N+3} x_{N+4} x_{N+1}$$

It is evident that any permutation combinator like $\pi_1$ can be constructed using only $B$ and $C$.

Assuming $c=m_{i,M}$, we get:

$$\varphi m_{i,M} s_1 s_2 = \pi_1 f_1 f_2 \ldots f_N \varphi c_T s_1 s_2 = c f_1 f_2 \ldots f_N s_1 s_2 \varphi = f_i s_1 s_2 \varphi$$

Let us assume that stack $s_1$ stores all of the non-blank tape squares to the left of the tape head but also contains the square immediately beneath the tape head at the top of the stack. The function $f$ corresponds to the action of the tape head when in state $i$. If the $i$-th state is a halt state, $f_i$ must return a pair of the two stacks and eliminate $\varphi$. Otherwise, the function $f$ must pop the stack $s_1$ and decide what to do next based on the result of the pop function.

$$f_i := B(BK)(BC(CI)) \qquad \textit{[if i is a halt state]}$$
$$\qquad BC(CI)(g_i)(g_i m_{1,M}(KI)) \qquad \textit{[if i is not a halt state]}$$

Recall that the pop funciton is `BC(CI)`, an empty stack is `(KI)`, and $m_{1,M}$ is set to be the blank character. So $f_i s_1$ returns $g_i m_{x,M} s_1^{-1}$, where $m_{x,M}$ is the content of the tape immediately beneath the tape head ($1 \leq x \leq M$) and $s_1^{-1}$ is the stack $s_1$ popped once. This works whether $s_1$ is empty or not. (One can avoid using two copies of $g_i$ by replacing $g_i$ with `BC(CI)` and appending $g_i$ at the end of the definition for $f_i$). Assuming the i-th state is not a halt state, we have:

$$\varphi m_{i,M} s_1 s_2 = f_i s_1 s_2 \varphi = g_i m_{x,M} s_1^{-1} s_2 \varphi$$

Now what must $g_i$ do at this step? It depends on what character was read as $m_{j,M}$. Since $m_{j,M}$ is itself a multiplexer function, $g_i$ can contain `M` functions $h_{i,1}$ to $h_{i,M}$ each corresponding to the appropriate action. $g_i$ can bring $m_{j,M}$ to the front and have it selects the appropriate function.

$$g_i := \pi_2 h_{i,1} h_{i,2} \ldots h_{i,M}$$

Such that,

$$\pi_2 x_1 x_2 \ldots x_M x_{M+1} = x_{M+1} x_1 x_2 \ldots x_M$$

Thus far we have:

$$\varphi m_{i,M} s_1 s_2 = f_i s_1 s_2 \varphi = g_i m_{x,M} s_1^{-1} s_2 \varphi = h_{i,x} s_1^{-1} s_2 \varphi$$

Let us define $y_{i,x}$ to be the character that must be written by the tape head when encountering character $x$ in state $i$. Let us also define $j_{i,x}$ to be the state that the tape head must transition to after encountering character $x$ in state $i$ and writing $y_{i,x}$.

To define $h_{i,x}$ we consider two cases for the direction that the tape head moves. If the tape must move to the left, $y_{i,x}$ must be pushed into $s_2$. If the tape moves to the right, $y_{i,x}$ must be pushed into $s_1^{-1}$ but then the top element of $s_2$ must be popped and also pushed into $s_1^{-1}$. For readability we will simply use $j$ instead of $j_{i,M}$.

$$h_{i,x} :=$$

$$B(CB(Py_{i,x}))(\pi_3 m_{j,M})$$ *[if tape head must move to the left]*

$$B(C(O(BC(B(B(\pi_3 m_{j,M}))P))(C(B(B(\pi_3 m_{j,M}))Pm_{1,M})(CK))))(Py_{i,x})$$ *[if tape head must move to the right]*

Where $m_{1,M}$ is the blank character and `P` and `O` are the push and pop functions (defined in section 4), respectively.

```
O := BC(CI)
P := BC(C(B(BK)))
```

I leave it to the reader to verify that $h_{i,x}$ followed by two stacks always results in $\pi_3 m_{j,M} s_1 (Py_{i,x} s_2)$ if the head must move to the left and $\pi_3 m_{j,M} (Pe(Py_{i,x} s_1)) s_2$ if the tape must move to the right (where e is the top element in $s_2$ – or the blank character if $s_2$ is empty). We can define $s_1^+$ $s_2^+$ as the state of the two stacks after this step. So far we have:

$$\varphi m_{i,M} s_1 s_2 = f_i s_1 s_2 \varphi = g_i m_{x,M} s_1^{-1} s_2 \varphi = h_{i,x} s_1^{-1} s_2 \varphi = \pi_3 m_{j,M} s_1^+ s_2^+ \varphi$$

Now all `q` must do is bring $\varphi$ to the front. So,

$\pi_3 := B(B(BC)C)(CI)$

$\pi_3 x_1 x_2 x_3 x_4 = x_4 x_1 x_2 x_3$

$\pi_3 m_{j,M} s_1^+ s_2^+ \varphi = \varphi m_{j,M} s_1^+ s_2^+$

Therefore we have $\varphi m_{i,M} s_1 s_2 = BC(CI) s_1 s_2$ if the state i is a halt state and $\varphi m_{i,M} s_1 s_2 = \varphi m_{j,M} s_1^+ s_2^+$ if i is not a halt state. We have fully implemented one iteration of a Turing machine. Notice that we have only used B, C, K, and I combinators. We did not need to use the W or M combinators that duplicate terms yet. However, φ was defined in terms of itself ($\varphi := \pi_1 f_1 f_2 \ldots f_N \varphi$). Similar to the previous recursive programs, φ can be redefined without self-reference using the Y combinator.

What this shows is that any computable function can be computed using combinators B, C, and K and self-reference. This is important because it proves that the third model in the manuscript (which is neither purely CL or λ-calculus) is universal. Furthermore, because there is a universal Turing machine that can also be simulated, and each iteration uses a fixed number of B, C, and K combinators, each iteration of the machine can be done in O(1) operations. So this system simulates Turing machines with linear complexity. In other words, any computable function can be computed with B, C, K, and a self-referencing system with identical algorithmic complexity to that of Turing machines.

## 8. A Brief Comparison of λ-calculus and Combinatory Logic

Despite there existing a one-to-one unique mapping between terms of λ-calculus and terms of CL, these systems differ at the level of basic operations. In λ-calculus there are no primitive functions. In CL there are no bounded variables. In some cases λ-calculus notations are easier to understand and compute functions in fewer elementary steps. For example, take the function that permutes 8 inputs in the following way:

**X**abcdefgh = debhgfca

In terms of primitive combinators, X must be expressed in an obscure expression, like the one below.

```
X = C(B(BC(B(BC)C))(C(B(B(B(B(BC)C))C)(B(B(B(B(B(B(BC)C)(B(B(BC)C)))C)(B(B(B(B(BC)C))C)))C
       … ))))))
```

Using this expression of X, it takes 94 operations to finish rearranging 8 inputs. But in λ-calculus the notation is far more simple and it complete with 10 straight-forward substitution operations.

The advantage of CL is does not require variable binding and the elementary operations may be simpler to physically implement. λ-calculus requires an to be an infinitely large set of variable names at its disposal. In the next section, I will discuss how RNA may serve as a vehicle for implementing either of these two computation systems or possibly even a hybrid system that takes advantage of features in both systems.

# 9. Summary

CL and λ-calculus are two equivalent computation system that produce an infinite set of functions that can be used to solve any computable problem. Much like modern day programming languages, it provides the tools for creative programming and there are unlimited ways to implement data structures and algorithms. What makes CL unique is that it is founded on a handful of very simply primitive combinators. To compute the output of function *f* on input *x*, it is sufficient to mindlessly execute of the leftmost combinator of the term *f(x)* until it is no longer possible to continue. The terms of λ-calculus can similarly be calculated by mindless iterative substitution/renaming operations, with the difference that in λ-calculus there are no primitive functions and the λ-notation is sufficient for determining what a function does.